\documentclass[a4paper, 11pt]{article}
\usepackage[english]{babel}
\usepackage{amsmath, amssymb, amsthm, amsfonts, a4wide, mathtools, hyperref}
\allowdisplaybreaks
\usepackage[caption=false]{subfig}
\usepackage[pdftex]{graphicx}
\usepackage{enumitem}
\usepackage{tikz}
\usetikzlibrary{arrows,positioning,calc}
\usepackage{multirow}
\usepackage{booktabs}
\usetikzlibrary{plotmarks}
\usetikzlibrary{patterns}
\tikzset{
    >=stealth',
    pt3/.style={
           rectangle,
           rounded corners,
           text width=10em,
           minimum height=3em,
           minimum width=6em,
           text centered},
            pt2/.style={
           rectangle,
           rounded corners,
           draw=blendedblue, thick,
           text width=6.5em,
           minimum height=2em,
           text centered}}

\newcommand{\nonu}{\nonumber \\[2mm]}
\newcommand{\be}{\begin{equation}}
\newcommand{\ee}{\end{equation}}
\newcommand{\bea}{\begin{eqnarray}}
\newcommand{\eea}{\end{eqnarray}}

\newcommand{\pp}{+ \!\!\! +}
\newcommand{\mm}{=}
\newcommand{\strutje}{\rule[-.2pt]{0in}{13pt}}

\newcommand{\tr}{\mathop{\text{tr}}}

\definecolor{blendedred}{rgb}{0.650,0.133,0.196}
\definecolor{blendedgreen}{rgb}{0,0.565,0.431}
\definecolor{blendedblue}{rgb}{0.102,0.388,0.612}
\definecolor{blendedyellow}{rgb}{0.843,0.725,0.294}
\definecolor{blendedorange}{rgb}{0.957,0.702,0.329}

\tikzset{line thickness/.style={line width=0.6pt}}

\newcommand{\circo}{\tikz[baseline=-0.5ex]{
    \draw [line thickness] (0,0.2mm) circle[radius=0.8mm];
}}

\newcommand{\circf}{\tikz[baseline=-0.5ex]{
    \draw [line thickness,fill] (0,0.2mm) circle[radius=0.8mm];
}}

\newcommand{\dash}{{\tikz[baseline=-0.5ex]{
      \useasboundingbox (-0.4mm,0) rectangle (+0.4mm,0.4mm);
      \draw [line thickness] (-0.5mm,0.2mm) -- (+0.5mm,0.2mm);
}}}

\date{\today}

\begin{document}
\title{Spinon bases in supersymmetric CFTs}
\author{Thessa Fokkema and Kareljan Schoutens \\[3mm]
{\it Institute for Theoretical Physics, University of Amsterdam} \\
{\it Science Park 904, 1098 XH Amsterdam, the Netherlands}}
%\institute{
%\inst{1} Institute for Theoretical Physics, University of Amsterdam  - Science Park 904, 1098 XH Amsterdam\\
%}
\maketitle

\begin{abstract}\small \noindent
We present a novel way to organise the finite size spectra of a class of conformal field theories (CFT) with $\mathcal{N}=2$ or (non-linear) $\mathcal{N}=4$ superconformal symmetry. Generalising the \emph{spinon basis} of the $SU(n)_1$ WZW theories, we introduce \emph{supersymmetric spinons} $(\phi^-, \phi^{+})$, which form a representation of the supersymmetry algebra. In each case, we show how to construct a multi-spinon basis of the chiral CFT spectra. The multi-spinon states are labelled by a collection $\{ n_j \}$ of (discrete) momenta. The state-content for given choice of $\{ n_j \}$ is determined through a generalised exclusion principle, similar to Haldane's `motif' rules for the $SU(n)_1$ theories. In the simplest case, which is the ${\cal N}=2$ superconformal theory with central charge $c=1$, we develop an algebraic framework similar to the Yangian symmetry of the $SU(2)_1$ theory. It includes an operator $H_2$, akin to a CFT Haldane-Shastry Hamiltonian, which is diagonalised by multi-spinon states. In all cases studied, we obtain finite partition sums by capping the spinon-momenta to some finite value. For the $\mathcal{N}=2$ superconformal CFTs, this \emph{finitisation} precisely leads to the so-called M$_k$ supersymmetric lattice models with characteristic order-$k$ exclusion rules on the lattice. Finitising the $c=2$ CFT with non-linear ${\cal N}=4$ superconformal symmetry similarly gives lattice model partition sums for spin-full fermions with on-site and nearest neighbour exclusion.
\end{abstract}

\section{Introduction and summary}
Conformal Field Theories (CFT) in two dimensions are highly structured thanks to the infinite dimensionality of the algebra of conformal transformations, which is (two copies of) the Virasoro algebra. Specific settings give rise to even larger algebras: superconformal and ${\cal W}$-algebras and, if a global symmetry group $G$ is respected, affine (or Kac-Moody) Lie algebras.  For rational CFTs exploiting conformal symmetry and its extensions leads to a formulation where the objects of interest (correlators, finite size spectra) are constructed as finite combinations of building blocks (conformal blocks, irreducible modules) whose structure is entirely dictated by the abundant symmetry.

There is an intimate relation between quantum critical lattice models in one spatial dimension (1D) and CFT, the latter describing the universal low-energy behaviour of the former. In specific examples, the relation extends to non-universal features and relates particular algebraic properties of the lattice model to the CFT. A particularly striking example is the algebraic structure of the $SU(n)$ Haldane-Shastry model and of the corresponding CFT, which has an $SU(n)_1$ affine (Kac-Moody) symmetry and central charge $c=n-1$~\cite{haldane:92, bernard:94, bouwknegt:94,bouwknegt:94b,schoutens:94au,bouwknegt:96}. This correspondence has led to a so-called spinon basis for chiral spectra of the CFT, and to a mathematical structure, called a Yangian quantum group, whose representation theory gives a generalised exclusion principle for the $SU(n)_1$ spinons. The Haldane-Shastry Hamiltonian has a CFT counterpart $H_2$, which can be explicitly diagonalised by multi-spinon states. Reading this correspondence backwards, one may recover the partition sums and symmetry operators of the Haldane-Shastry lattice models through a procedure called finitisation, which reduces the CFT partition sum to a finite lattice model partition sum. 

The idea of spinon-bases for rational CFTs, and finitisations thereof, has been explored in more general terms, see for example~\cite{schoutens:97xh,bouwknegt:98}. 
\begin{table}
  \begin{center}
    \begin{tabular}{cc}
      \toprule
      qH state & CFT \\
      \midrule
      \strutje RR$_k$ state at $\nu={k \over kM+2}$ & $SU(2)_{k,M}$ WZW theory at $c={3 \over k+2}$ \\[1mm]
      \strutje NASS$_k$ state at $\nu={2k \over 2kM+3}$ & $SU(3)_{k,M}$ WZW theory at $c={8 \over k+3}$ \\[1mm]
      \bottomrule
    \end{tabular}
  \end{center}
  \caption{CFT data for the Read-Rezayi (RR) and non-Abelian spin singlet (NASS) quantum Hall states. The integer $M$ is a deformation parameter in the CFT; even (odd) $M$ describe bosonic (fermionic) states. Both the RR$_k$ and NASS$_k$ states are characterised by an order-$k$ clustering of the fundamental particles, which are spin-polarised for the RR states and carry spin-$\frac{1}{2}$ for the NASS states.}
  \label{tab:qhcft}
\end{table}

Before turning to the supersymmetric CFTs that are the main subject of this paper, we comment on another important condensed matter connection of particular CFTs, which is the so-called CFT-quantum Hall (CFT-qH) correspondence~\cite{moore:91}. This correspondence relates qH wavefunctions (ground states and quasi-hole excitations) to conformal blocks in specific CFTs. Exploiting this correspondence has led to the definition of two important series of non-Abelian qH states: the Read-Rezayi (RR$_k$) states for spin-less fermions with order-$k$ clustering~\cite{read:98} and the non-Abelian spin-singlet (NASS$_k$) states~\cite{ardonne:99}. The CFTs going with the RR$_k$ states are deformations (set by an integer $M$) of the $SU(2)_k$ WZW theory, while the NASS$_k$ states correspond to deformations of the $SU(3)_k$ WZW theory, see Table~\ref{tab:qhcft}. The paper~\cite{ardonne:01} applied finitised partition sums for parafermionic CFTs to the solution of the counting problem of quasi-hole excitations over Read-Rezayi and NASS states.

In the simple case of the $SU(2)_1$ CFT, the triple relation
\begin{equation*}
\begin{tikzpicture}
 \node[pt3] (lattice) at (-2,0) {Lattice model};
 \node[pt3] (cft) at (2.75,0) {quantum Hall state};
 \node[pt3] (qh) at (0,1.8) {CFT};
 \draw[to-to, line width=0.8pt] (-1.3,0.7)--(-0.7, 1.3);
 \draw[to-to, line width=0.8pt] (0.7, 1.3)--(1.3, 0.7);
 \draw[to-to, line width=0.8pt] (-0.4,0)--(0.4,0);
\end{tikzpicture}
\end{equation*}
can be extended in all directions, as the ground state wave function of the $SU(2)$ Haldane-Shastry lattice model precisely takes the form of a Jastrow factor \mbox{$\prod_{i<j} (z_i-z_j)^2$}, which is nothing else than the $\nu={1 \over 2}$ bosonic Laughlin (or $k=1$, $M=0$ RR) qH wave function. References~\cite{greiter:11, paredes:12, nielsen:11} presented attempts to generalise this structure to the $M=0$ bosonic RR$_k$ states at higher $k$, by constructing parent Hamiltonians which have the $k>1$ RR$_k$ states as their exact ground states. The papers~\cite{tu:13, glasser:15} pursued a similar program for the cases $(k=1,M=q-2)$ and $(k=2,M=q-1)$, which are the $\nu={1 \over q}$ Laughlin and Moore-Read states, both in a 1D and 2D setting. The lattice models that have come out of these various constructions are less structured than the $SU(2)$ Haldane-Shastry model, and they show no signs of special symmetries or integrability~\cite{thomale:11}.

In this paper we follow a different approach and focus on the finitisation of the $M=1$ (fermionic) CFTs in Table~\ref{tab:qhcft}. In this we are guided by the supersymmetry that these CFTs possess. For the RR$_k$ series this supersymmetry, which is what remains of the $SU(2)$ symmetry after the deformation $M=0\rightarrow M=1$, takes the form of ${\cal N}=2$ superconformal symmetry, while for the  NASS$_k$ theories the remnant of the $SU(3)_1$ symmetry is a non-linear ${\cal N}=4$ superconformal algebra, which includes an $SU(2)_1$ subalgebra.

Our main focus (sections~\ref{spinonbasisk1}--\ref{sec:n=1k=1spinon}) will be on the simplest case, which is the $M=1$ deformation of the $SU(2)_1$ CFT, and which we recognise as the simplest unitary minimal model of ${\cal N}=2$ superconformal symmetry at central charge $c=1$. We introduce supersymmetric spinons and construct a multi-spinon basis of the chiral CFT spectrum. This construction entails a generalised exclusion principle for supersymmetric spinons. Carrying through the finitisation leads to a supersymmetric lattice model, known as the $M_1$ model, which was first introduced in \cite{fendley:03_2}. The basic supersymmetry algebra, which involves a $U(1)$ (fermion number) operator $J_0$, supercharges $Q^\pm_0$ and hamiltonian $H_1$, reads
\begin{equation}
  [J_0,Q_0^\pm] = \pm Q_0^\pm, \qquad \{ Q_0^+,Q_0^-\} = 2H_1, \qquad [H_1, Q_0^\pm] = 0 .
\end{equation}
It is explicitly realised in both the CFT and the $M_1$ lattice models. Tracing through the finitisation, we identify supersymmetric ground states $|\psi_G\rangle$ satisfying
\begin{equation}
  Q_0^+ |\psi_G\rangle = 0, \qquad Q_0^- |\psi_G\rangle = 0, \qquad H_1 |\psi_G\rangle = 0
\end{equation}
in the $M_1$ lattice model on open and closed chains. These results have been confirmed by independent analysis using Bethe ansatz and cohomology techniques~\cite{fendley:03,fendley:05_2}.

In the $k=1$, $M=1$ CFT, we identify higher symmetry operators, such as $Q_1^+$ and $H_2$, and show that the `Hamiltonian' $H_2$ is explicitly diagonalised by multi-spinon states. In the $SU(2)_1$ case, such operators have direct counterparts on the lattice. In the supersymmetric case, we have been unable to identify lattice operators that are tractable and reflect the CFT higher symmetry structure directly on the lattice.

In section~\ref{sec:n=2kg1} we turn to the $M=1$ deformations of the $SU(2)_k$ WZW theories, which form the minimal series of ${\cal N}=2$ superconformal field theory. 
For $k>1$ the fusion product of two supersymmetric spinons has more than a single channel, which complicates the construction of a multi-spinon basis. The corresponding problem for $SU(2)_k$ has been analysed and solved in~\cite{bouwknegt:94c}. Here we focus on the case $M=1$, $k=2$, where we establish a multi-spinon basis and determine finitised partition sums. They correspond to the so-called supersymmetric M$_2$ model, first defined in~\cite{fendley:03}. In a similar fashion, finitising the $k$-th minimal model of ${\cal N}=2$ superconformal field theory leads to the M$_k$ models of \cite{fendley:03}. Configurations in the M$_k$ models satisfy an exclusion rule stating that at the most $k$ consecutive sites can be occupied. This rule is similar to the $k$-clustering condition in the RR$_k$ states; both have their origin in the $Z_k$ parafermion fields that are part of the $SU(2)_{k,M}$ CFTs.

In our final section~\ref{sec:N=4spin}, we turn to $SU(3)_{k,M}$ CFTs corresponding to the NASS$_k$ qH states. We focus on the simplest fermionic case, with $k=1$ and $M=1$, which is a $c=2$ CFT. The deformation of the $SU(3)_1$ affine Kac-Moody algebra is a superconformal algebra with bosonic currents for spin and charge degrees of freedom (forming a $U(1)\times SU(2)_1$ affine algebra) and ${\cal N}=4$ supercurrents which carry spin ${1 \over 2}$ as well as unit charge. The algebra is different from the linear ${\cal N}=4$ superconformal algebras that are known in the literature - one distinguishing feature is the presence of non-linear terms in the defining OPEs of two of the supercurrents. Despite these complications we have managed to construct once again a supersymmetric spinon basis and to give an exclusion rule specifying the state-content of a multi-spinon word. Finitising this basis leads to well-structured finite partition sums. They can be matched with configurations of spin-${1 \over 2}$ fermions on open chains, subject to the constraint that each site has at most one fermion, and that nearest neighbour sites cannot be simultaneously occupied.

\section{Spinon-basis for ${\cal N}=2$, $k=1$ minimal model}\label{spinonbasisk1}
\subsection{Superconformal currents and basis supersymmetry algebra}
In this first minimal model, the currents pertaining to the ${\cal N}=2$ superconformal symmetry have the following expressions in terms of a chiral boson \cite{waterson:86}
\be
T(z)=-{1 \over 2}\partial \varphi \partial \varphi, \quad G^\pm(z) = \sqrt{2 \over 3} e^{\pm i \sqrt{3} \varphi}, \quad  J(z)= {i \over \sqrt{3}} \partial \varphi.
\ee 
The associated current-modes  
\be
L_n =\oint {dz \over 2\pi i} z^{n+1}, \quad  G^\pm_s = \oint {dz \over 2\pi i} z^{s+1/2} G^\pm(z), \quad J_m = \oint {dz \over 2\pi i} z^m J(z)
 \ee
satisfy the (anti-)commutation relations of the ${\cal N}=2$ superconformal algebra
\begin{eqnarray}
&& \{ G_r^\pm, G_s^\mp \} = {c \over 3}(r^2-{1 \over 4})\delta_{r+s} + 2L_{r+s}  \pm (r-s) J_{r+s} 
\nonu
&& [L_m, G^\pm_r] = ({m \over 2} - r) G_{m+r}^\pm, \qquad [J_m, G^\pm_r] = \pm G_{m+r}^\pm
\nonu
&& [L_m,L_n]= {c \over 12}m(m^2-1)\delta_{m+n} + (m-n) L_{m+n}  
\nonu
&& [L_m,J_n] = -nJ_{m+n}, \qquad [J_m,J_n] = {c \over 3}\delta_{m+n}
\end{eqnarray}
with $c=1$. In the Neveu-Schwarz (NS) sector, where $l\in Z+{1 \over 2}$, the basic supersymmetry algebra is
\be
[J_0,Q_0^\pm] = \pm Q_0^\pm, \qquad \{ Q_0^+,Q_0^-\} = 2H_1, \qquad [H_1, Q_0^\pm] = 0 
\label{eq:susyalg1}
\ee
with
\be 
Q_0^+=G_{-1/2}^+, \quad Q_0^-=G_{1/2}^-, \quad H_1 = L_0 - {1 \over 2} J_0 \ .
\label{eq:susyalgNS}
\ee
In the Ramond (R) sector, where $l\in Z$, we have the same algebra eq.~\eqref{eq:susyalg1}, but with
\be 
Q_0^+=G_0^+, \quad Q_0^-=G_0^-, \quad H_1 = L_0 - {c \over 24} \ .
\label{eq:susyalgR}
\ee
Our goal here is to organise the finite size spectra of this CFT by employing spinon operators and higher symmetries (other than the conformal currents), in close analogy to the description of the $SU(2)_1$ CFT with the help of spinons and Yangian symmetry \cite{haldane:92}.  We work this out for the NS spectra - the R spectra are connected to these through spectral flow and show a similar structure.

\subsection{Higher symmetry operators}
Working in the NS sector, we define a higher supercharge
\begin{equation}
  \label{eq:q1p}
  \begin{aligned}
    Q_1^+ & = \oint {dz_1 \over 2\pi i} \oint {dz_2 \over 2\pi i} 
    \left[ {z_2\over z_1-z_2} J(z_1)G^+(z_2) - {z_1\over z_1-z_2} G^+(z_1)J(z_2) \right]
    \\
    & = \sum_{l\geq 0} \left(  J_{-l-1} G^+_{l+1/2} - G^+_{-l-1/2} J_l \right) .
  \end{aligned}
\end{equation}
In this expression we assume radial ordering, $|z_1|>|z_2|$. Anti-commuting this new supercharge with $Q_0^-$ defines a Haldane-Shastry type Hamiltonian in the CFT
\begin{equation}
     \label{eq:h2p}
  \begin{aligned}
  H_2 &\equiv \{ Q_1^+, Q_0^- \}
  \\
  &=  J_0J_0 
  + 2 \sum_{l\geq 0} \left(  J_{-l-1} L_{l+1} - L_{-l} J_l  \right) 
  \\ &
  + \sum_{l\geq 0} \left(  G^+_{-l-1/2} G^-_{l+1/2} + G^-_{-l-1/2} G^+_{l+1/2} + 2(l+1) J_{-l-1} J_{l+1}   \right)  .
\end{aligned}
\end{equation}
Note that $H_2$ is not hermitian. By construction
\begin{equation}
  [ H_1, Q_1^+] = 0, \qquad [H_2, Q_0^-]=0 \ .
\end{equation}

\subsection{Supersymmetric spinons}
\label{sec:spbase}
\begin{figure}
  \centering
   \includegraphics{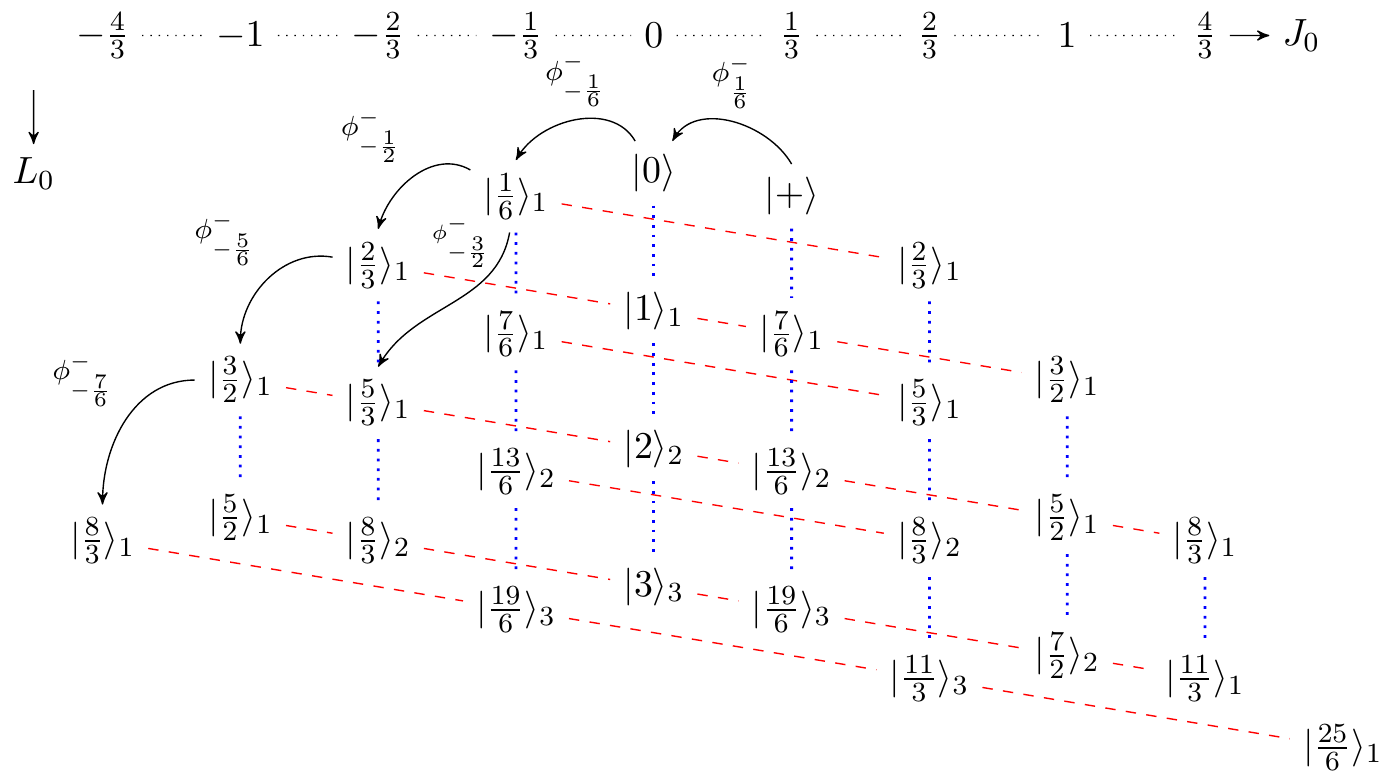}
  \caption{Chiral spectrum of the $\mathcal{N}=2$ supersymmetric CFT at $c=1$. The states are denoted by $|h\rangle_d$, where $d$ is the degeneracy of the state. The supercharges $Q_0^{\pm}$ act parallel to the red dashed lines. The black arrows show the construction of the fully polarized multi-spinon states. The blue dotted lines connect states with the same charge, the Virasoro generators act along these lines. \label{fig:cftsu2wzwsusy}}
\end{figure}

The Virasoro primary fields 
\be
\phi^-(z) = e^{-i \frac{1}{\sqrt{3}} \varphi}(z), \qquad \phi^{+}(z) = e^{2i \frac{1}{\sqrt{3}} \varphi}(z) \ .
\ee
form a doublet under supersymmetry
\begin{equation}
\begin{aligned}
& [ G^+_{s-1/2}, \phi^-_{-1/6-n} ] = \sqrt{2 \over 3} \phi^{+}_{-2/3+s-n}\\
& \{ G^-_{s+1/2}, \phi^{+}_{-2/3-n} \} = \sqrt{2 \over 3} (3n-2s+1) \phi^{-}_{-1/6+s-n} .
\end{aligned}
\end{equation}
Using the modes of these {\it supersymmetric spinons} we can write the following general  {\it multi-spinon states}
\be
\Phi^{\{\alpha_j\}}_{\{n_i\}} =
\phi^{\alpha_N}_{(3-2N)/6-s_N-n_N} \ldots \phi^{\alpha_2}_{-1/6-s_2-n_2} \phi^{\alpha_1}_{1/6-s_1-n_1}  |+\rangle
\label{eq:multiphi}
\ee
with $\alpha_i=-,+$ and $s_i=0$ if $\alpha_i=-$ and $s_i=1/2$ if $\alpha=+$. The $\{n_i\}$ are integers satisfying
$n_N\geq n_{N-1}\geq \ldots n_1\geq 0$. Furthermore,
\be
| + \rangle, \qquad |0\rangle \propto \phi^-_{1/6} |+\rangle
\ee
are supersymmetry singlets satisfying 
\be
H_1  |+\rangle =0, \quad H_2  |+\rangle =0, \quad H_1  |0\rangle =0, \quad H_2  |0\rangle =0 .
\ee
All other multi-spinon states in eq.~\eqref{eq:multiphi} are part of a supersymmetry doublet. We refer to Figure~\ref{fig:cftsu2wzwsusy} for a graphical representation.

Our choice of taking $|+\rangle$ rather than $|0\rangle$ as the reference state in~\eqref{eq:multiphi} has been made to guarantee that all states in the CFT finite size spectrum are covered by the multi-spinon states. However, the states~\eqref{eq:multiphi} are not linearly independent. In subsection \ref{sec:spinonbasis}  below we explain how a complete set of linearly independent multi-spinon states, {\it i.e.}, a {\it multi-spinon basis}, can be constructed. 

\subsection{One-spinon states}
On 1-spinon states
\be
\Phi^-_n = \phi^-_{1/6-n} |+\rangle, \qquad \Phi^{+}_n = \phi^{+}_{-1/3-n} |+\rangle,
\ee
the supersymmetry charges act as
\begin{equation}
  Q_0^+ \Phi^-_n= \sqrt{2 \over 3} \Phi^{+}_n, \quad Q_0^- \Phi^{+}_n= \sqrt{2 \over 3} (3n) \Phi^-_n,
  \label{eq:q0phi1}
\end{equation}
while $Q_1^+$ acts as 
\begin{equation}
  Q_1^+ \Phi^-_n= \sqrt{2 \over 3} (n-{2 \over 3})  \Phi^{+}_n.
  \label{eq:q1ph1}
\end{equation}
As a consequence
\begin{equation}
  H_1 \Phi^\alpha_n = n \Phi_n^\alpha, 
  \quad 
  H_2 \Phi^\alpha_n= 2n (n-{2 \over 3})  \Phi_n^\alpha.
\end{equation}
These explicit evaluations require expressions for some normal ordered field products, which can be extracted from the BPZ descendant series, adapted to the present situation. Details are provided in appendix \ref{sec:appA}.

We remark that the state $\Phi_{n=0}^{+}$ vanishes identically, in accordance with the fact that $\Phi_{n=0}^{-}=|0\rangle$ is a supersymmetry singlet. 
The dimension of the space generated by supersymmetric spinons with mode index $n_1$ becomes
\begin{equation}
  \label{eq:dim1spinon}
  d[n_1] = 
  \begin{cases} 
    2 & \text{if}\ n_1>0 , \\
    1 & \text{if}\ n_1=0 .
  \end{cases}
\end{equation}
The vanishing of $\Phi_{n=0}^{+}$ is a first example of the generalised exclusion principle for supersymmetric spinons, which we present in subsections~\ref{sec:exclusion} and~\ref{sec:spinonbasis} below.

\subsection{Two spinons: $H_2$ eigenstates}
The 2-spinon states take the form
\begin{equation}
  \begin{aligned}
    & \Phi^{-,-}_{n_2,n_1} =\phi^-_{-1/6-n_2} \phi^-_{1/6-n_1} |+\rangle, \quad  &&\Phi^{+,-}_{n_2,n_1} =\phi^{+}_{-2/3-n_2} \phi^-_{1/6-n_1} |+\rangle\\
    & \Phi^{-,+}_{n_2,n_1} =\phi^-_{-1/6-n_2} \phi^{+}_{-1/3-n_1} |+\rangle, \quad &&\Phi^{+,+}_{n_2,n_1} =\phi^{+}_{-2/3-n_2} \phi^{+}_{-1/3-n_1} |+\rangle.
  \end{aligned}
\end{equation}
They all share the $H_1$ eigenvalue
\begin{equation}
  h_1[n_1,n_2] = n_1+n_2+{1 \over 3} .
\end{equation}
In appendix~\ref{sec:appB} we provide explicit expressions for the action of $Q_1^+$ and $H_2$ on these states. 
It turns out that $H_2$ is `upper triangular' in the sense that $H_2$ acting on a state $\Phi^{\alpha_2,\alpha_1}_{n_2,n_1}$ leads to a combination of states with labels $(n_2',n_1')$ of the form $(n_2-l,n_1+l)$, with $l\geq 0$. This implies that $H_2$ can be diagonalised with relative ease. Explicit expressions are provided in appendix~\ref{sec:appB}. The eigenvalues and eigenstates are found to be
\begin{equation}
  \begin{aligned}
    [n_2,n_1;\alpha_1=-], \quad H_2=h_2[n_2,n_1]: &  \quad \Psi^{-,-}_{n_2,n_1}, \quad \Psi^{+,-}_{n_2,n_1}
    \\
    [n_2,n_1;\alpha_1=+], \quad H_2=h^\prime_2[n_2,n_1]: &  \quad \Psi^{-,+}_{n_2,n_1}, \quad  \Psi^{+,+}_{n_2,n_1},
  \end{aligned}
\end{equation}
with
\begin{equation}
  \label{eq:h2eigenvals}
  \begin{aligned}
    h_2[n_2,n_1] &= 2(n_2+{1 \over 3})^2 + 2n_1(n_1-{2 \over 3}) \\
    h^\prime_2[n_2,n_1] &= 2(n_2+{1 \over 3})(n_2-{5 \over 3}) + 2n_1(n_1-{2 \over 3}) .
  \end{aligned}
\end{equation}

\subsection{Two spinons: algebraic structure}
\begin{figure}
  \centering
  \includegraphics{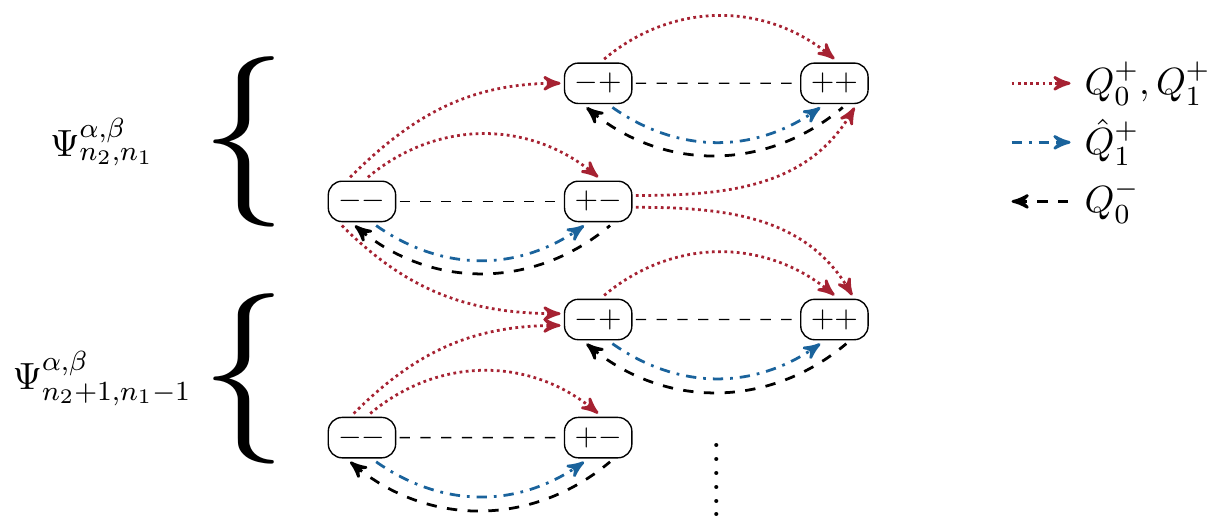}
  \caption{Action of various supercharges on the eigenstates $\Psi_{n_2,n_1}^{\alpha_2,\alpha_1}$ of $H_2$. $Q_0^-$ and $\widehat{Q}_1^+$ act within the eigenstate doublets, while $Q_0^+$ and $Q_1^+$ act as ladder operators, connecting doublets with different $H_2$ eigenvalues.}
    \label{fig:multiplets}
\end{figure}

The supercharge $Q_0^-$ acts on the $H_2$ eigenstates at $H_1$ eigenvalue $h_1[n_2, n_1]=n_2+n_1+{1 \over 3}$ according to
\begin{equation}
  \begin{aligned}
    \sqrt{\frac{3}{2}} Q_0^- \Psi^{+,+}_{n_2,n_1} &= (3n_2+1) \Psi^{-,+}_{n_2,n_1}\\
    \sqrt{\frac{3}{2}} Q_0^- \Psi^{+,-}_{n_2,n_1} &= (3n_2+1)\Psi^{-,-}_{n_2,n_1}\\
    \sqrt{\frac{3}{2}} Q_0^- \Psi^{-,+}_{n_2,n_1} &= 0.\\
  \end{aligned}
\end{equation}
The action on these same states of $Q_0^+$ and $Q_1^+$ is more difficult to obtain. These operators do not commute with $H_2$ and they are thus expected to mix states with different $H_2$ eigenvalues. Remarkably, this mixing is limited to two terms only: the multiplet $[n_2,n_1;\alpha_1=-]$ gets mapped into $[n_2,n_1;\alpha_1=+]$ and $[n_2+1,n_1-1;\alpha_1=+]$. Explicitly,
\begin{equation}
  \begin{aligned}
    \sqrt{\frac{3}{2}} Q_0^+ \Psi^{-,-}_{n_2,n_1} &=  3 {h_1[n_2,n_1] \over 3n_2+1} \Psi^{+,-}_{n_2,n_1} + \Psi^{-,+}_{n_2,n_1} +\lambda_{n_2,n_1} \Psi^{-,+}_{n_2+1,n_1-1}\\
    \sqrt{\frac{3}{2}}Q_0^+ \Psi^{+,-}_{n_2,n_1} &= - \Psi^{+,+}_{n_2,n_1} -\lambda_{n_2,n_1} \frac{1+3n_2}{4+3n_2} \Psi^{+,+}_{n_2+1,n_1-1}\\
    \sqrt{\frac{3}{2}}Q_0^+ \Psi^{-,+}_{n_2,n_1} &= 3 {h_1[n_2,n_1] \over 3n_2+1} \Psi^{+,+}_{n_2,n_1}
  \end{aligned}
  \label{eq:actionQ0+}
\end{equation}
and
\begin{equation}
  \begin{aligned}
    \sqrt{\frac{3}{2}} Q_1^+ \Psi^{-,-}_{n_2,n_1} &=  {3 \over 2} {h_2[n_2,n_1] \over 3n_2+1} \Psi^{+,-}_{n_2,n_1} + (n_1-{2 \over 3}) \Psi^{-,+}_{n_2,n_1} +n_2 \lambda_{n_2,n_1} \Psi^{-,+}_{n_2+1,n_1-1}\\
    \sqrt{\frac{3}{2}}Q_1^+ \Psi^{+,-}_{n_2,n_1} &= - (n_1-{2 \over 3}) \Psi^{+,+}_{n_2,n_1} - n_2 \lambda_{n_2,n_1} \frac{1+3n_2}{4+3n_2} \Psi^{+,+}_{n_2+1,n_1-1}\\
    \sqrt{\frac{3}{2}}Q_1^+ \Psi^{-,+}_{n_2,n_1} &= {3 \over 2} {h^\prime_2[n_2,n_1] \over 3n_2+1} \Psi^{+,+}_{n_2,n_1},
  \end{aligned}
  \label{eq:actionQ1+}
\end{equation}
with 
\begin{equation}
  \lambda_{n_2,n_1}=\frac{3 (3n_2+4)(3n_2-3n_1+3)(3n_2-3n_1+4)}{(3n_1-1)(3n_2-3n_1+2)(3n_2-3n_1+5)} \ .
\end{equation}
The explicit action of the operators $Q_0^+$ and $Q_1^+$ on 2-spinon states inspires the definitions
\begin{equation}
  \widehat{Q}^+_1 \equiv Q^+_1 - (H_1-{2 \over 3})Q_0^+ + {1 \over 4}[Q_0^+, H_2], \quad \widehat{H}_2 \equiv \{Q_0^-,\widehat{Q}_1^+\}.
  \label{eq:hatQ1}
\end{equation}
Acting on 2-spinon states, $\widehat{Q}^+_1$ has the properties
\begin{equation}
  (\widehat{Q}^+_1)^2=0, \quad [\widehat{Q}^+_1,H_2]=0,
\end{equation}
and we have the explicit action
\begin{equation}
\begin{aligned}
  & \widehat{Q}^+_1 \Psi^{-,-}_{n_2,n_1}= \sqrt{2 \over 3}(-2n_1+{2 \over 3})\Psi^{+,-}_{n_2,n_1}, \quad \widehat{Q}^+_1 \Psi^{+,-}_{n_2,n_1} = 0\\
  & \widehat{H}_2 \Psi^{\pm,-}_{n_2,n_1} = \hat{h}_2[n_2,n_1] \Psi^{\pm,-},   \quad \hat{h}_2[n_2,n_1]=-{4 \over 9}(3n_2+1)(3n_1-1)\\
  & \widehat{Q}^+_1 \Psi^{-,+}_{n_2,n_1}= \sqrt{2 \over 3}(-2n_1-{4 \over 3})\Psi^{+,+}_{n_2,n_1}\\
  & \widehat{H}_2 \Psi^{\pm,+}_{n_2,n_1} = \hat{h}^\prime_2[n_2,n_1]\Psi^{\pm,+}_{n_2,n_1},  \quad \hat{h}^\prime_2[n_2,n_1] =-{4 \over 9}(3n_2+1)(3n_1+2).
\end{aligned}
\end{equation}
With all these operators in place we have a neat organisation of the collection of all 2-spinon states. They organize into doublets $[n_2,n_1;\alpha_1]$ that are characterised by their $H_2$ (or, equivalently, $\widehat{H}_2$) eigenvalues. The supercharges $Q_0^-$ and $\widehat{Q}_1^+$ act within these doublets, while $Q_0^+$ and $Q_1^+$ act as ladder operators connecting the $[n_2,n_1;\alpha_1=-]$ doublet to $[n_2,n_1;\alpha_1=+]$ and $[n_2+1,n_1-1;\alpha_1=+]$. This structure is depicted in Figure~\ref{fig:multiplets}.

The one issue that remains to be addressed is that of null-states among the 2-spinon eigenstates that we have listed.

\subsection{Exclusion principle and dimension formula for 2-spinon states}
\label{sec:exclusion}
For general mode indices $\{n_i\}$, the multi-spinon states of equation~\eqref{eq:multiphi} represent $2^N$ independent states, which give rise to $2^{N-1}$ doublets under supersymmetry. It is important to realise that for special choices of the $\{n_i\}$ some of these state are actually vanishing.  The `vanishing rules' for supersymmetric spinons will be closely analogous to the generalised exclusion principle satisfied by $SU(2)$ spinons in the $SU(2)_1$ WZW CFT.

A simple example is the 1-spinon state with $\alpha_1=-$ and $n_1=0$. This state (which is proportional to the CFT vacuum $|0\rangle$) is a supersymmetry singlet and the would-be superpartner with $\alpha_1=+$ and $n_1=0$ vanishes.

Another example where a reduction in the number of states is observed are the two-spinon states with $n_1=0$ (2 states rather than 4). Furthermore, it turns out that for $n_2=n_1>0$, the states 
$\Psi^{-,+}_{n_2,n_1}$ and $\Psi^{+,+}_{n_2,n_1}$ vanish.  A simple example is $\Psi^{+,+}_{1,1}$. This state has $J_0={5 \over 3}$ and $L_0={19 \over 6}$, whereas all non-vanishing states in the CFT spectrum 
with $J_0={5 \over 3}$ have $L_0\geq {25 \over 6}$.

To demonstrate the vanishing of the doublet $[n_2=n_1;\alpha_1=+]$ for general $n_1\geq 1$ we resort to the generalised commutation relations (g.c.r.) satisfied by modes of $\phi^+(w)$. Using the OPE
\begin{equation}
  \phi^+(z)\phi^+(w) = (z-w)^{4 \over 3}\left[ \phi^{(4)}(w) + \ldots \right], 
  \qquad 
  \phi^{(4)}(z) = e^{i \frac{4}{\sqrt{3}} \varphi}(z)
\end{equation}
one derives the g.c.r. (see for example~\cite{fateev:85, bouwknegt:94b})
\begin{equation}
  \sum_{l\geq 0} C_l^{-{4 \over 3}} \left[ \phi^+_{-{2 \over 3}-n-l} \phi^+_{-{1 \over 3}-m+l} |+\rangle - \phi^+_{-{5 \over 3}-m-l} \phi^+_{+{2 \over 3}-n+l} |+\rangle \right] =0,
\end{equation}
where 
\begin{equation}
  (1-x)^{a} = \sum_{l\geq0} C_l^a x^l, \quad C_l^a = (-1)^l \left( \begin{array}{c} a \\ l \end{array} \right).
\end{equation}
Choosing $m=n=n_1$, comparing with the explicit expression for $\Psi_{n_2,n_1}^{+,+}$ in equation~\eqref{eq:psialphaplus},
\begin{equation}
  \Psi^{+,+}_{n_1,n_1} =  \sum_{l\geq 0} \rho_l \, \phi^+_{-{2 \over 3}-n_1-l} \phi^+_{-{1 \over 3}-n_1+l} |+\rangle,
\end{equation}
with $\rho_0=1$ and $\rho_{l>0}$ given in equation~\eqref{eq:etazetarho}, and using
\begin{eqnarray}
  \rho_l &=&  \binom{l-{1 \over 3}}{l} {1 \over 2} {4 \over 5} \ldots {3l-2 \over 3l-1} =\binom{ l-{2 \over 3}}{l}= (-1)^l \binom{-{1 \over 3}}{l}  \nonu
         &=& (-1)^l \binom{-{4 \over 3}}{l} - (-1)^{l-1} \binom{-{4 \over 3}}{l-1}=  C_l^{-{4 \over 3}} - C_{l-1}^{-{4 \over 3}},
\end{eqnarray}
we find that indeed the state $\Psi_{n_1,n_1}^{+,+}$ vanishes.

We conclude that the space spanned by 2-spinon states $\Psi^{\alpha_2,\alpha_1}_{n_2,n_1}$ has dimension
\begin{equation}
  \label{eq:dim2spinon}
  d[n_2,n_1] =
  \begin{cases}
    4 & \text{if}\ n_2 > n_1>0 , \\
    2 & \text{otherwise}.
  \end{cases}
\end{equation}
This rule will be generalized to general multi-spinon states in the next subsection. 

\subsection{Multi-spinon states as a basis of the CFT}
\label{sec:spinonbasis}
We now consider the general multi-spinon state, as written in equation~\eqref{eq:multiphi}. The eigenvalues of $J_0$ and $H_1$ on these states are
\begin{eqnarray}
  && j_0[(n_N,\alpha_N), \ldots, (n_1,\alpha_1)]= {1-N \over 3} + 2(s_N+s_{N-1}+\ldots + s_1)
     \nonu
  &&  h_1[(n_N,\alpha_N), \ldots, (n_1,\alpha_1)]= \sum_{j=1}^N (n_j+ {j-1 \over 3}) .
\end{eqnarray}
We claim that the general expression for the $H_2$ eigenvalues is
\begin{equation}
  \begin{aligned}
    h_2&[(n_N,\alpha_N), \ldots, (n_1,\alpha_1)]=\sum_{j=1}^N \left( 2\left(n_j +\frac{j-1}{3}\right)^2 \right.\\
    &\left. - 4 (n_j+{j-1 \over 3})( {2-j \over 3}+2(s_{j-1}+s_{j-2}+\ldots + s_1)) \right).
  \end{aligned}
  \label{eq:generalh2}
\end{equation}
Our final task in specifying the basis is to point out which of the $2^N$ eigenstates are non-vanishing. Starting from the expression~\eqref{eq:dim2spinon} for $N=2$, we claim that the dimension incurs an extra factor of 2 upon adding an additional $\phi$-mode with label $n_j$ if and only if $n_j>n_{j-1}$. The precise statement is that a doublet of $H_2$ eigenstates, written as (for $N\geq 2$)
\begin{equation}
  [n_N,n_{N-1},\ldots n_1;\alpha_{N-1},\ldots, \alpha_1]
\end{equation}
is vanishing for $\alpha_j=+$ as soon as $n_{j+1}=n_j$ or, for $j=1$, as $n_1=0$.

The general dimension formula for a state with $N\geq 2$ spinons becomes
\begin{equation}
  d[n_N,n_{N-1},\ldots,n_1] = \prod_{j=3}^N  2^{\epsilon_j} d[n_2,n_1],
  \label{eq:dimNspinon}
\end{equation}
with
\begin{equation}
  \epsilon_j = 
  \begin{cases} 
    1 & \text{if}\ n_j>n_{j-1}
    \\
    0 & \text{if}\ n_j=n_{j-1} ,
  \end{cases}
\end{equation}
and with $d[n_2,n_1]$ as in eq.~\eqref{eq:dim2spinon}.

Important evidence for this claim comes from inspecting the CFT character formulas that are implied by these dimension formulas. In section~\ref{sec:characters} below we work this out and demonstrate  that the characters of the CFT are correctly reproduced.

\subsection{Algebraic structure}
The analysis of the algebraic structure of operators acting on $N$-spinon states is straightforward but cumbersome. As for the case $N=2$, the operators $Q_0^+$ and $Q_1^+$ will act as ladder operators between doublets with different $H_2$ eigenvalues. We expect that the definition~\eqref{eq:hatQ1} of an operator $\widehat{Q}_1^+$ commuting with $H_2$ can be extended to the $n$-spinon sector, but have not found a closed form expression.

We remark that the simple structure of the $H_2$ eigenvalues suggests that the $N$-spinon sector is in essence a tensor product of 1-spinon states,
\begin{equation}
  \Psi^{\alpha_N,\ldots \alpha_2,\alpha_1}_{n_N,\ldots n_2,n_1} \quad \leftrightarrow \quad \phi^{\alpha_N}_{(n_N+\frac{N-1}{3})} \ldots \phi^{\alpha_2}_{(n_2+{1 \over 3})} \otimes \phi^{\alpha_1}_{(n_1)} \otimes |+\rangle.
\end{equation}
In this notation, the actions of $Q_0^-$, $H_1$ and $H_2$ are completely described by their action on the reference state,
\begin{equation}
  Q_0^-|+\rangle=0, \quad H_1|+\rangle=0, \quad H_2 |+\rangle=0
\end{equation}
and on 1-spinon states
\begin{equation}
  Q_0^- \phi^{+}_{(n)} = \sqrt{2 \over 3} \, 3n \, \phi^-_{(n)}, \quad  
  H_1 \phi^{\alpha}_{(n)} = h_1[n]  \phi^{\alpha}_{(n)}, \quad
  H_2 \phi^{\alpha}_{(n)} = h_2[n] \phi^\alpha_{(n)},
\end{equation}
with
\begin{equation}
  h_1[n] = n, \quad h_2[n] = 2(n+{1 \over 3})^2,
\end{equation}
and by the co-products
\begin{eqnarray}
  \Delta(Q_0^-) &=& Q_0^- \otimes 1
                    \nonu
                    \Delta(H_1) &=& H_1 \otimes 1 + 1 \otimes H_1
                                    \nonu
                                    \Delta(H_2) &=& H_2 \otimes 1 + 1 \otimes H_2 - 4 \, H_1 \otimes J_0.
                                                    \label{eq:copr}
\end{eqnarray}
The action of the operators $Q_0^+$ and $Q_1^+$ (such as given in eq.~\eqref{eq:actionQ0+} and~\eqref{eq:actionQ1+}) cannot be captured through a simple co-product rule. We expect though that on a general $N$-spinon state further operators, such as an $N$-spinon generalisation of $\widehat{Q}_1^+$, can be defined that do admit a simple characterisation in terms of a closed form co-product. The resulting algebraic structure takes the place of the Yangian that organises the CFT spectrum of the $SU(2)_1$ theory.

\section{Finitisation and CFT characters}
We proceed by explaining the procedure of finitisation of the CFT finite size spectra. To this end, we will cap the integer $n_N$ that encodes the highest spinon momentum in an $N$-spinon state~\eqref{eq:multiphi} according to
\begin{equation}
  n_N +{2N -3 \over 6} = {2L-1 \over 6} - l,
  \label{eq:fink1}
\end{equation}
with $l$ a non-negative integer.
In the example $L=5$ this bound allows a 6-spinon state with $n_6=n_5=\ldots=n_1=0$, 3-spinon states with $1 \geq n_3 \geq n_2 \geq n_1 \geq 0$ and the reference state $|+\rangle$. We list all states in Table~\ref{table:spec}, where we include the eigenvalues of $J_0$, $H_1$ and $H_2$. 

It will turn out that the finitisation eq.~\eqref{eq:fink1}  leads to a finite partition sum that can be associated to a supersymmetric lattice model on an open chain with $L$ sites. Sending $L$ to infinity, and keeping track of the lattice model partition sum, will reproduce the characters of the CFT finite size spectra.

Before we come to that we introduce an alternative labelling of the multi-spinon states in terms of \emph{rapidities}.
\begin{table}
  \centering
    \begin{tabular}{lllll}
      \toprule
      $n_i$ & $J_0$ & $H_1$ & $H_2$ & $ \{ m_j \} $ \\[1mm]
      \midrule 
      - & 1/3 & 0 & 0 & \{ 1, 4, 7, \ldots \} \\
      000 & - 2/3 & 1 & 2 & \{ 4, 7, \ldots \} \\
      100 & - 2/3 & 2 & 8, 4/3 & \{ 3, 7, \ldots \} \\
      110 & - 2/3 & 3 & 34/3 & \{ 2, 7, \ldots \} \\
      111 & - 2/3 & 4 & 12 & \{ 1, 7, \ldots \} \\
      000000 & - 5/3 & 5 & 30 & \{ - , 7, \ldots\} \\
      \bottomrule
    \end{tabular}
  \caption{Multi-spinon states corresponding to the truncation to $L=5$ and the corresponding description in terms of rapidities $\{m_i\}$}
  \label{table:spec}
\end{table} 

\subsection{Rapidity labelling} 
We label multi-spinon states in terms of integers (rapidities) $\{m_i \geq 1\}$ satisfying 
\begin{equation}
  m_{i+1} \ge m_i+3, \qquad i=1,2,\ldots
\end{equation}
revealing a type of exclusion statistics that is closely analogous to the statistics displayed by the $SU(2)_1$ theory.  We show below that allowing a finite range for the $m_i$,
\begin{equation}
  1 \leq m_i \leq L-1
\end{equation}
is equivalent to capping the highest spinon momentum, as in eq.~\eqref{eq:fink1}. We proceed in close analogy to the similar analysis in the $SU(n)_1$ CFTs~\cite{haldane:92}.

The systematics of assigning rapidities to multi-spinon states are as follows. The reference state $|+\rangle$ corresponds to $\{m_i^0\} = \{1,4,7,\ldots\}$; excited states with $J_0 = 1/3-k$ are obtained by taking out $m_1, m_2, \ldots m_k$ and performing shifts on the remaining $\{m_j\}$. The $J_0$ and $H_1$ eigenvalues are
\begin{equation}
  J_0 = 1/3-k, \qquad H_1 = \sum_i [m_i^0-m_i] \ .
\end{equation}
The leading $H_2$ eigenvalue $h_2$ is given by
\begin{equation}
  h_2 = 
  2k(k-1)(k+2)
  + \sum_i  [ P_2(m_i^0) - P_2(m_i) ]
  -4 \sum_i (i-1)(3k+3i-2-m_i)
  \label{eq:evh2}
\end{equation}
with
\begin{equation}
  P_2(m) = {4 \over 3} m^2 - {10 \over 3} m + 4 \ .
\end{equation}
The second term in eq.~\eqref{eq:evh2} is a sum of 1-particle energies, whereas the third term has its origin in the non-trivial co-product of $H_2$. 

To recover the spinon modes $n_1,n_2,\ldots$ from a sequence $\{m_i\}$ we define increments
\begin{equation}
\begin{aligned}
  \lambda_0&=m_1-1 \\
  \lambda_j &= m_{j+1}-m_j-3 & \text{for}\ &  j\geq 1.
\end{aligned}
\end{equation} 
The spinon modes are then obtained as
\begin{equation}
  n_1 n_2 \ldots n_N = \underbrace{00 \ldots 0}_{\lambda_0} \underbrace{11 \ldots 1}_{\lambda_1} \ldots  \underbrace{tt \ldots t}_{\lambda_t} \ .
\end{equation}

We can now specify the states that survive the finitisation to the range $1 \leq m_i \leq L-1$, with $L=3l-1$. Adding a tail $\{3l+1,3l+4,\ldots\}$ to the rapidity sequence brings us back to the CFT, where we can identify all corresponding multi-spinon states. One easily checks that this precisely selects all $N$-spinon states with 
\begin{equation}
  n_N +{2N -3 \over 6} \leq {2L-1 \over 6}, \qquad N=0,3,\ldots, 3l \ .
\end{equation}

To complete our description we need to specify sectors where the number of spinons is 1 or 2 modulo 3, and to capture the shifts of the eigenvalues of the operator $H_2$ within a set of multi-spinon states with given $\{n_j\}$.

The CFT vacuum $|0\rangle$, which in our conventions is a 1-spinon state, corresponds to the sequence $\{ 2,5,8, \ldots\}$. Taking out $k$ of the $m_i$ leads to $(k+1)$-spinon states with
\begin{equation}
  J_0 = -k, \qquad H_1 = \sum_i [m_i^0-m_i] \ .
\end{equation}
The leading $H_2$ eigenvalue $h_2$ is given by
\begin{equation}
  h_2 = 
  2k(k^2+2k-1)
  + \sum_i  [ P_2(m_i^0) - P_2(m_i) ]
  -4 \sum_i (i-1)(3k+3i-1-m_i) \ .
\end{equation}
The finite size $L$ is now of the form $L=3l$, selecting $N$ spinon states with
\begin{equation}
  n_N +{2N -3 \over 6} \leq {2L-1 \over 6}, \qquad N=1,4,\ldots, 3l+1\ .
\end{equation}
The sequence $\{ 3,6,9, \ldots\}$ corresponds to the supersymmetry doublet
\begin{equation}
  \Psi^{-,-}_{0,0}, \quad \Psi^{+,-}_{0,0}
\end{equation}
with 
\begin{equation}
  h_1[0,0]=1/3, \quad h_2[0,0]=2/9 \ .
\end{equation}
Taking out $k$ of the $m_i$ leads to $(k+2)$-spinon states with
\begin{equation}
  J_0 = -{1 \over 3} - k, \qquad H_1 = {1 \over 3} + \sum_i [m_i^0-m_i] \ .
\end{equation}
The leading $H_2$ eigenvalue $h_2$ is given by
\begin{equation}
  h_2 = 
  2 k^3 +6 k^2+{4 \over 3} k + {2 \over 9}
  + \sum_i  [ P_2(m_i^0) - P_2(m_i) ]
  -4 \sum_i (i-1)(3k+3i-m_i) \ .
\end{equation}
The finite size $L$ is now of the form $L=3l+1$, selecting $N$ spinon states with
\begin{equation}
  n_N +{2N -3 \over 6} \leq {2L-1 \over 6}, \qquad N=2,5,\ldots, 3l+2 \ .
\end{equation}
We finally describe the shifts in the $H_2$ eigenvalues within the collection of states with given $\{n_j\}$. For each $i>0$ such that $\lambda_i>0$ we have that
\begin{equation}
  n_{j+1}-n_j > 0, \quad j= \sum_{i'<i} \lambda_{i'} = m_i-3i+2 \ . 
\end{equation}
Assuming $j>0$, subtraction of the co-product terms for $s_j=+$ and $s_j=-$ leads to a shift $\delta_i h_2$ (using $n_{j+1}=i$)
\begin{equation}
  \delta_i h_2 = - 4 {m_i+2 \over 3} \Theta[m_i>3i-2] .
\end{equation}
The total shift becomes
\begin{equation}
  \delta h_2 = -4 \sum^*_{i>0} \Theta[m_{i+1} > m_i+3] \Theta[m_i>3i-2]  {m_i+2 \over 3} \delta_i \ ,
  \label{eq:shift}
\end{equation}
where * indicates the absence of the term where
\begin{equation}
  m_1=2 \quad \text{and} \quad m_i=3i-1\ .
\end{equation}
For each term in eq.~\eqref{eq:shift} there is a choice $\delta_i =0,1$, leading to a number of (generally different) $H_2$ eigenvalues. Obviously, the information encoded through the labelling with $\{m_i\}$ is equivalent to that in equation~\eqref{eq:generalh2} and the dimension formulas~\eqref{eq:dim1spinon},~\eqref{eq:dim2spinon} and~\eqref{eq:dimNspinon}. Comparing the two we identify
\begin{equation}
  \delta_i = 2 \, s_{m_i-3i+2} \ .
\end{equation}

\subsection{Character formulas}
\label{sec:characters}
In the previous subsections we described the multi-spinon CFT states that survive finitisation with parameter $L$. They give rise to a partition function
\begin{equation}
  Q_L(w,q) = \tr [w^{3J_0} q^{H_1}] \ .
\end{equation}
Examples for some small sizes are
\begin{equation*}
\begin{aligned}
   Q_0 &= 1\\
  Q_1 &= q^{1 \over 3} w^{-1}[1+ w^3]\\
   Q_2 &= q w^{-2} [1 + w^3 (1 + q^{-1})]\\
   Q_3 &= q^2 w^{-3} [1+w^3 (1 + q^{-1} + q^{-2}) + w^6 q^{-1}]\\
   Q_4 &= q^{10 \over 3} w^{-4} [1+w^3 (1 + q^{-1} + q^{-2} + q^{-3} ) + w^6 (q^{-1} + q^{-2} + q^{-3})] \\
   Q_5 &= q^5 w^{-5}[1+ w^3 (1 + q^{-1} + q^{-2} + q^{-3} + q^{-4}) \\
  & \qquad \qquad \ + w^6 (q^{-1} + q^{-2} + 2q^{-3} + q^{-4} + q^{-5}) + w^9 q^{-3}] .
  \end{aligned}
\end{equation*}
The dimension formulas~\eqref{eq:dim1spinon},~\eqref{eq:dim2spinon} and~\eqref{eq:dimNspinon} translate into a recursion relation obeyed by the $Q_L$,  
\begin{equation}
  Q_L = Q_{L-3} + q^{L \over 3} w^{-1}[Q_{L-1} + w^3 Q_{L-4}] \ .
  \label{eq:recur}
\end{equation}
This relation implies that for $w=1$, $q=1$ the $Q_L$ are Fibonacci numbers. 

Using the recursion relation~\eqref{eq:recur} we find the following general expression
\begin{equation}
  \begin{split}
    Q_L(w,q)& = \sum_{m \equiv {L+1} (\text{mod }3)} w^{1-m} q^{\frac{m^2 -m}{6}} \binom{\frac{2L+2 +m}{3}}{\frac{L+1-m}{3}}_q \\
    &=\sum_{p=0,1,2,\ldots}^{\left \lfloor{(L+1)/2}\right \rfloor} w^{3p-L} q^{(L- 3p)(L-3p+1) \over 6}  \binom{L+1-p}{p}_q,
  \end{split}
\end{equation}
where $\binom{a}{b}_q$ is the $q$-binomial. 
Sending $L\to \infty$ leads to
\begin{equation}
  \underset{\lim l \rightarrow \infty}{Q_{3l+k}}= \sum_{m \equiv -k\ \text{mod}\ 3} w^m q^{m(m-1) \over 6}  \prod_{l \geq 1} \frac{1}{1-q^l} \ .
\end{equation}
This expression correctly reproduces the well-known affine $U(1)$ characters in the CFT. [Note that the leading $q$-power for given $m$ is set by the eigenvalue of $H_1=L_0-{1 \over 2}J_0$, where $L_0={m^2 \over 6}$ and $J_0={m \over 3}$.] We take this result as strong evidence for the correctness of the generalised exclusion principle for supersymmetric spinons, as specified in section~\ref{sec:spinonbasis}.

\section{${\cal N}=2$, $k=1$ Supersymmetric lattice model}
\label{sec:n=1k=1spinon}
The algebraic structure going with the supersymmetric spinon basis of the supersymmetric CFT is closely analogous to that of the $SU(2)$ spinon basis and the Yangian symmetry of the $SU(2)_1$ CFT. In the latter case, the rapidity labelling of multi-spinon states in the CFT is given by
\be
m_{i+1} \ge m_i+2, \qquad i=1,2,\ldots \ .
\ee
The CFT vacuum corresponds to $\{ 1,3,5,\ldots \} $, while the state labelled as $\{ 2,4,6,\ldots \}$ is the lowest 1-spinon state with conformal dimension $h={1 \over 4}$. The $SU(2)$ multiplicities in the Yangian multiplet labelled by a set $\{ m_i \}$ can be extracted through Haldane's {\it motif} rules \cite{haldane:92}. Allowing a finite range for the $m_i$,
\be
1 \leq m_i \leq L-1
\ee
leads to finite partition sums, which are easily recognized as describing a collection of $L$ spin-${1 \over 2}$ spins. Remarkably, the same {\it motif} rules that organize the CFT spectrum, when applied to the finitised $L$-spin states, precisely reproduce the degeneracies of the celebrated Haldane-Shastry hamiltonian for spins on $L$ sites. In this correspondence, the $L$ spins are positioned at locations $z_j=\exp {{2\pi i \over L}j}$, $j=0,1,\ldots,L-1$ in the complex plane and the Haldane-Shastry Hamiltonian is given by
\be
H_{\rm HS} = \sum_{i \neq j} {z_i z_j \over z_{ij} z_{ji} } [P_{ij} -1] \ ,
\ee
where $z_{ij}=z_i-z_j$ and $P_{ij}$ permutes the spins on locations $z_i$ and $z_j$. We now investigate to what extent the CFT-to-lattice model correspondence of the Haldane-Shastry model generalizes to the fermionic case where $SU(2)$ spin symmetry has been replaced by ${\cal N}=2$ superconformal symmetry. 

\subsection{Spin-less fermions with nearest neighbour exclusion}
We remark that the partition sum $Q_L(w, q=1) $ precisely matches that of a lattice model of spin-less fermions on an open, $L$-site chain \emph{with exclusion of nearest neighbour occupation}, with the fermion counting operator $f$ given by
\begin{equation}
  f=J_0+{L \over 3}.
\end{equation}
For the example $L=3$, the configurations are
\begin{equation*}
\begin{aligned}
  f=0  \quad & \circo\dash\circo\dash\circo \\
          f=1  \quad& \circf \dash\circo\dash\circo, \circo\dash\circf\dash\circo, \circo\dash\circo\dash\circf\\
                  f=2 \quad & \circf\dash\circo\dash\circf.
\end{aligned}
\end{equation*} 

In addition, the sum of $Q_{L-1}(w, q=1)$ and $Q_{L-3}(w, q=1)$ matches the state counting of the same model on a closed chain with $L$ sites. Keeping track of the eigenvalues of the translation operator $T$ we have the following result for the closed chain partition sum $Z_{L}(w,q)$  
\begin{equation}
  \begin{aligned}
    Z_{L} (w,q)  & =  \tr\left( w^{3F} q^{\frac{N}{2\pi i} \log{T}} \right)\\
    &= q^{-{L(L-1) \over 6}} w^{L-1} \left[ Q_{L-1} + w^3 q^{-L+1} Q_{L-3} \right],
  \end{aligned}
  \label{eq:closedchainpf}
\end{equation}
where we used $q^L=1$ to establish the correspondence. 

In second quantized formalism, the constraint of excluding nearest neighbour occupation can be incorporated by defining dressed creation and annihilation operators
\begin{equation}
  d_i = \prod_{<ij>} c_i (1-n_j), \qquad d^\dagger_i = \prod_{<ij>} c^\dagger_i (1-n_j) \ ,
  \label{eq:dressedc}
\end{equation}
where $<ij>$ denotes the nearest neighbour relation and $n_j=c^\dagger_j c_j$. 

In first quantized formalism, with lattice sites $w_j=\omega_L^j$, $\omega_L=\exp{2 \pi i \over L}$, the constraint, together with the fermi statistics of the particles, leads to wavefunctions of the general form
\begin{equation}
  \Psi(z_1, \ldots, z_n) = \prod_{i<j} (z_i-z_j)(z_i-\omega_Lz_j)(\omega_Lz_i-z_j) P(z_1, \ldots, z_n) \ ,
\end{equation}
with $P(z_1,\ldots,z_n)$ a fully symmetric polynomial. In the large-$L$ limit, where $\omega_L \to 1$, the wavefunction develops triple zero's, establishing a close connection with the Laughlin wavefunction for a $1/3$ filled lowest Landau level. 

\subsection{Lattice model supercharges and Hamiltonian $H_1$}
\label{sec:level0}
In the lattice model we define  
\begin{equation}
  J_0 = \sum_i d^\dagger_i d_i, \qquad Q_0^-= \sum_i d_i, \qquad Q_0^+  = \sum_i d^\dagger_i 
\end{equation}
and 
\begin{equation}
  2 H_1 = \{ Q_0^-, Q_0^+ \} \ . 
\end{equation}
These definitions can be made on any graph specified through a set of vertices and nearest neighbour relations. The resulting `basic supersymmetric lattice model' or `M$_1$ model' was first introduced by Fendley, de Boer and one of the present authors in \cite{fendley:03}. It has been studied extensively in 1D and on higher dimensional lattices, see for example \cite{fendley:03_2,beccaria:04,fendley:05_2,huijse:08,huijse:11}.

A direct application of the $M_1$ lattice model-to-CFT correspondence is the determination of the ground states on an $L$-site open chain. The chiral CFT has two supersymmetric ground states at $H_1=0$,
\begin{equation}
  |+\rangle, \quad |0\rangle
\end{equation}
in agreement with the fact that the Witten index is $W=2$. The state $|+\rangle$ is part of $Q_L$ with $L\equiv 2 \mod 3$ and leads to a supersymmetric lattice model ground state with $f={L+1 \over 3}$. Similarly, the CFT ground state $|0\rangle$ is part of $Q_L$ with $L\equiv 0 \mod 3$ and leads to a supersymmetric lattice model ground state with $f={L \over 3}$.

For the $L$-site chain with periodic boundary conditions the partition sum is expressed in the characters $Q_{L-1}$ and $Q_{L-3}$ through equation~\eqref{eq:closedchainpf}. This leads us to the following ground states for the closed chain lattice model with $L$ sites 
\begin{center}
  \begin{tabular}{ll}
    $L=3l$: & two supersymmetric ground states at $f=l$ \\[2mm]
    $L=3l-1$: & one supersymmetric ground state at $f=l$ \\[2mm]
    $L=3l+1$: & one supersymmetric ground state at $f=l$.
  \end{tabular}
\end{center}

\subsection{Higher symmetry operators on the lattice}
The analogy with the Haldane-Shastry model, where the algebraic structure of Yangian symmetry and higher (Haldane-Shastry) conserved quantities is shared between the CFT and the lattice model, invites the investigation of operators such as $Q_1^+$, $H_2$, $\widehat{Q}_1^+$ and $\widehat{H_2}$ in the lattice model. We have investigated such operators, but have been unable to identify operators that are tractable and reflect the CFT higher symmetry structure directly on the lattice. 

\section{${\cal N}=2$ superconformal field theory with $k>1$}\label{sec:n=2kg1}
We now turn to the extension of the ideas to $k$-th minimal models of ${\cal N}=2$ superconformal field theory. Through the CFT-qH connection, these correspond to the simplest $M=1$ fermionic Read-Rezayi (RR$_k$) states at filling $\nu={k \over k+2}$. In their bosonic guise, the RR$_k$ states are characterised by an order-$k$ clustering: the wave function vanishes when $k$ or more particles come to the same position. This property can be traced to the presence of $Z_k$ parafermions in the CFT. Our main result in this part of the paper will be that a very similar property of order-$k$ clustering arises in the lattice models that are obtained by finitising the same CFTs. This clustering property, stating that at most $k$ neighbouring sites on the lattice can be simultaneously occupied by spin-less fermions, is at the basis of the definition of the so-called supersymmetric M$_k$ models, which were first introduced in~\cite{fendley:03} and further studied in~\cite{hagendorf:14, fokkema:15, hagendorf:15b}. For the sake of clarity, we restrict our presentation to the case $k=2$.  

\subsection{$k=2$ minimal model of ${\cal N}=2$ superconformal field theory}
The $k=2$ minimal model, at central charge $c_2={3 \over 2}$, is built from a scalar field $\varphi$ plus Ising fields $\psi$, $\sigma$. The ${\cal N}=2$ supercurrents take the form
\begin{equation}
  G^\pm= \sqrt{1 \over 2} \psi e^{\pm i \sqrt{2} \varphi}
\end{equation}
and the $U(1)$ current is $J(z)={i \over \sqrt{2}}\partial \varphi$. The basic supersymmetry algebra is again given by eqs.~\eqref{eq:susyalg1} and~\eqref{eq:susyalgNS}. Our definition of  spinon fields combines a scalar field vertex operator with the Ising sector spin-field $\sigma$, 
\begin{equation}
\begin{aligned}
  \phi^-&= \sigma \, e^{-i {\sqrt{2} \over 4} \varphi} :&& J_0=-{1 \over 4}, & h={1 \over 8}\\
  \phi^+&= \sigma \, e^{3i {\sqrt{2} \over 4} \varphi} : && J_0={3 \over 4}, & h={5 \over 8}.
\end{aligned}
\end{equation}
The reference state for the spinon basis is the state created from the vacuum by $e^{i {\sqrt{2} \over 2} \varphi}$, which we write as $|{1 \over 2}\rangle$. The chiral NS spectrum has three supersymmetry singlets with $H_1=0$, 
\begin{equation}
  |0\rangle = \phi^-_{1 \over 16} \phi^-_{3 \over 16}  |\tfrac{1}{2}\rangle, \qquad
  \phi^-_{3 \over 16} |\tfrac{1}{2}\rangle, \qquad |\tfrac{1}{2}\rangle.
\end{equation}

\subsection{Fusion channels and spinon basis}
Before writing multi-spinon states, we remark that the spinon fields are chiral vertex operators (CVO) that intertwine Ising sectors $1$, $\psi$ and $\sigma$, according to
\begin{equation}
  \phi: 1 \rightarrow \sigma, \quad \phi: \sigma  \rightarrow 1+\psi, \quad \phi: \psi \rightarrow \sigma .
\end{equation}
An $N$-spinon state thus involves a fusion path $1\to \tau_1 \to \tau_2 \ldots$ with $\tau_j=1,\sigma,\psi$. In a general multi-spinon state
\begin{equation}
  \phi^-_{{1 \over 8}-\Delta_N-n_N} \ldots \phi^-_{{1 \over 8}-\Delta_2-n_2} \phi^-_{{1 \over 8}-\Delta_1-n_1}  |\tfrac{1}{2}\rangle
\end{equation}
the minimal mode sequence, corresponding to $n_j=0$ for all $j$, is set by increments $\Delta_j$. In~\cite{schoutens:98} these were found to be given by $\Delta_1=0$ and
\begin{equation}
  \Delta_{j+1} = 
  \begin{cases}
    \Delta_j &  \text{if $\tau_{j+1}\tau_j\tau_{j-1} = 1\sigma 1$ or $\psi \sigma \psi$}
    \\
    \Delta_j+{1 \over 4}  & \text{if $\tau_{j+1}\tau_j\tau_{j-1} = \sigma 1\sigma$ or $\sigma \psi \sigma$}
    \\
    \Delta_j +{1 \over 2} & \text{if $\tau_{j+1}\tau_j\tau_{j-1} = 1\sigma \psi$ or  $\psi \sigma 1$}.
  \end{cases}
\end{equation}
  As before, we need to determine the state-content of a general $N$-spinon state with $n_N\geq \ldots \geq n_2\geq n_1\geq 0$. The result is in essence the same as for the $k=1$ case, 
equation~\eqref{eq:dim1spinon},~\eqref{eq:dim2spinon} and~\eqref{eq:dimNspinon}, with a small modification due to the presence of three instead of two supersymmetry singlets. The dimension formula becomes
\begin{equation}
  d[n_N,n_{N-1},\ldots,n_1] = \prod_{j=2}^N  2^{\epsilon_j} d[n_1]
  \label{eq:dimNspinonk2}
\end{equation}
with
\begin{equation}
  \epsilon_j = 
  \begin{cases}
    1 & \text{if  $n_j>n_{j-1}$ or $[n_3=0, \tau_2=1]$ or $[n_2=0, \tau_2=\psi]$} 
    \\
    0 & \text{otherwise},
  \end{cases}
\end{equation}
and with $d[n_1]$ as in equation~\eqref{eq:dim1spinon}.

We checked numerically that this prescription precisely reproduces the chiral characters of this supersymmetric CFT.  To illustrate the result, let us focus on the CFT states with $J_0=-{1 \over 2}$. The leading contributions to the CFT character $\chi_{\text{CFT}}=\tr(q^{L_0})$ in this sector come from 4-spinon states, according to (we denote the fusion channel by $\tau_4\tau_3\tau_2\tau_1$)
\begin{align*}
  1 \sigma 1 \sigma : &\quad \phi_{-{1 \over 8}-n_4} \phi_{-{1 \over 8}-n_3} \phi_{{1 \over 8}-n_2} \phi_{{1 \over 8}-n_1} && \quad { q^{1 \over 4} \over (q)_4}\\
  1 \sigma \psi \sigma: &\quad \phi_{-{9 \over 8}-n_4} \phi_{-{5 \over 8}-n_3} \phi_{-{3 \over 8}-n_2} \phi_{{1 \over 8}-n_1} && \quad { q^{9 \over 4} \over (q)_4}\\
  \psi \sigma 1 \sigma : &\quad \phi_{-{5 \over 8}-n_4} \phi_{-{1 \over 8}-n_3} \phi_{{1 \over 8}-n_2} \phi_{{1 \over 8}-n_1} && \quad { q^{3 \over 4} \over (q)_4}\\                                                                                                                                                                                                                                                                   \psi \sigma \psi \sigma: &\quad \phi_{-{5 \over 8}-n_4} \phi_{-{5 \over 8}-n_3} \phi_{-{3 \over 8}-n_2} \phi_{{1 \over 8}-n_1} && \quad { q^{7 \over 4} \over (q)_4}.
  \end{align*}
This results in
\begin{equation}
  \begin{aligned}
    \chi_{4-\text{sp}}(q) &= { q^{1 \over 4} +  q^{3 \over 4} +  q^{7 \over 4} +  q^{9 \over 4} \over (q)_4}\\
    & = q^{1 \over 4}[1 + q^{1 \over 2} + q + 2q^{3 \over 2} + 3q^2 + 3q^{5 \over 2} + 4q^3 + \ldots ]\ .
  \end{aligned}
  \end{equation}

Comparing with the full character in this sector
\begin{equation}
  \begin{aligned}
    \chi_{\text{CFT}} &=q^{1 \over 4} {\prod_{l \geq 0} (1+q^{l+{1 \over 2}}) \over \prod_{l \geq 1} (1-q^l)} \\
    &= q^{1 \over 4}[1 + q^{1 \over 2} + q + 2q^{3 \over 2} + 3q^2 + 4q^{5 \over 2} + 5q^3 + \ldots]
  \end{aligned}
\end{equation}
we observe that the lowest discrepancy is a missing state at $q^{11 \over 4}$. This state is filled in as the superpartner of the lowest 8-spinon state in fusion channel $1 \sigma 1 \sigma 1 \sigma 1 \sigma 1$, at $L_0={9 \over 4}$.

\subsection{Finitisation and M$_2$ lattice models}
Having established the spinon basis, it is an easy step to finitise the spectrum by capping spinon momenta. We restrict to states with an even number of spinons and cap the momenta according to 
\begin{equation}
  n_N -{1 \over 8} + \Delta_N = {2L-1 \over 8} - l,
  \label{eq:fink2}
\end{equation}
with $l$ a non-negative integer.

We work out some small sizes and list the contributions to the partition sums $Q_L=\tr(w^{4J_0}q^{H_1})$ in Table~\ref{tab:finitespin}.
\begin{table}
\centering
\begin{tabular}{lrrl}\toprule
Length & fusion channel & $n_j$ & contribution\\[1mm]\midrule
  $L=0$ & $1\sigma$ & $00$ & $q^0$\\[2mm]
$L=1$ & $1\sigma 1 \sigma$ & $0000$ & $w^{-2}q^{1 \over 2}(1+w^4)$\\[4pt]
 $L=2$ & $\psi \sigma$ & $00$ & $w^0 q^{1 \over 2}(1+w^4)$\\
  & $ 1\sigma 1 \sigma 1 \sigma $ & $ 000000 $ & $ w^{-4}q^{3 \over 2}(1+w^4)$ \\[4pt]
$L=3 $ & &$ - $ & $ w^2$\\
   & $ \psi \sigma 1 \sigma $ & $ 0000 $ & $ w^{-2}q(1+w^4)$\\
   & $ \psi \sigma \psi \sigma $ & $ 0000 $ & $ w^{-2}q^2(1+w^4)$\\
   & $ 1\sigma 1 \sigma 1 \sigma 1 \sigma $ & $ 00000000 $ & $ w^{-6}q^3(1+w^4)$\\[4pt]
$ L=4 $ & $ 1\sigma $ & $ 00 $ & $ q^0$\\
   & $ 1\sigma  $ & $ 10 $ & $ q^1(1+w^4)$\\
   & $ 1\sigma $ & $ 11 $ & $ q^2(1+w^4)$\\
   & $ \psi \sigma 1 \sigma 1 \sigma $ & $ 000000 $ & $ w^{-4}q^2(1+w^4)$ \\
   & $ \psi \sigma \psi \sigma 1 \sigma  $ & $ 000000 $ & $ w^{-4}q^3(1+w^4)$ \\
   & $ \psi \sigma \psi \sigma \psi \sigma  $ & $ 000000 $ & $ w^{-4}q^4(1+w^4)$ \\
   & $ 1 \sigma 1\sigma 1 \sigma 1 \sigma 1 \sigma $ & $ 0000000000 $ & $ w^{-8}q^5(1+w^4)$ \\
  \bottomrule
\end{tabular}
\caption{Contributions to the finitised partitions sums $Q_L$ in the $\mathcal{N}=2$ supersymmetric CFT with $k=2$.}
\label{tab:finitespin}
\end{table}
Collecting all the terms we find,
\begin{equation*}
\begin{aligned}
  Q_0 =& 1\\
  Q_1 =& q^{1 \over 2} w^{-2} [1+ w^4] \\
   Q_2 =& q^{3 \over 2} w^{-4} [1 + w^4 (q^0 + q^{-1})+ w^8 q^{-1}] \\
   Q_3 =& q^3 w^{-6} [1+w^4 (q^0 + q^{-1} + q^{-2}) + w^8 (q^{-1}+q^{-2}+q^{-3})]\\
   Q_4 =& q^5 w^{-8} [1+w^4 ( q^{0} + q^{-1} + q^{-2} + q^{-3} )\\
  & \qquad \qquad \ + w^8 (q^{-1} + q^{-2} + 2q^{-3}+q^{-4}+q^{-5})+w^{12}(q^{-3}+q^{-4})].
\end{aligned}
\end{equation*}
Putting $q=1$ we observe that the partition sum $Q_L(w,q=1)$ now matches that of a lattice models of spin-less fermions on an open, $L$-site chain \emph{with exclusion rule stating that at most $k=2$ consecutive lattice sites can be occupied}. In this, the fermion counting operator $f$ takes the value
\begin{equation}
  f = J_0+{L \over 2}.
\end{equation}
For the example $L=3$ the configurations are 
\begin{equation*}
\begin{aligned}
  f=0 \quad & \circo \dash \circo \dash\circo \\
 f=1 \quad&   \circf \dash\circo\dash\circo, \circo\dash\circf\dash\circo, \circo\dash\circo\dash\circf\\
 f=2 \quad & \circf \dash \circf \dash \circo, \circf \dash \circo \dash \circf, \circo \dash \circf \dash \circf .
\end{aligned}
\end{equation*}
In reference~\cite{fendley:03} this precise lattice model configuration space was taken as the basis of the definition of the $M_2$ lattice model. As for the M$_1$ model, the finitisation procedure immediately points at supersymmetric ground states of the M$_2$ model on the open chain. The CFT ground state $|{1 \over 2}\rangle$ is part of $Q_L$ with $L\equiv 3 \mod 4$ and leads to a supersymmetric lattice model ground state with $f={L+1 \over 2}$. Similarly, the CFT ground state $|0\rangle$ is part of $Q_L$ with $L\equiv 0 \mod 4$ and leads to a supersymmetric lattice model ground state with $f={L \over 2}$.

We end with a comment on the finitised characters with an odd number of spinons. It is easily checked that those correspond to partition sums of the $M_2$ model on open chains with $\sigma$-type boundary conditions, as introduced and discussed in \cite{fokkema:15,hagendorf:15b}. The proper procedure is now to cap the spinon momentum as
\begin{equation}
  n_N -{1 \over 8} + \Delta_N = {2L-1 \over 8} - l,
  \label{eq:fink2prime}
\end{equation}
where $l$ is now non-negative \emph{integer or half-integer}. This leads to partition sums $Q^\prime_L$, such as
\begin{equation*}
\begin{aligned}
   Q^\prime_0 &= w\\
   Q^\prime_1 &= q^{1 \over 4} w^{-1} [1+ w^4] \\
   Q^\prime_2 &= q^1 w^{-3} [1 + w^4 (q^0 + q^{-1})] \\
   Q^\prime_3 &= q^{9 \over 4} w^{-5} [1+w^4 (q^0 + q^{-1} + q^{-2}) + w^8 (q^{-1}+q^{-2})].
  \end{aligned}
  \end{equation*}
They give a perfect match with the M$_2$ model on an open $L$-site chain, with, on one end, a $\sigma$-type boundary condition forbidding the simultaneous occupation of the 2 sites closest to the boundary \cite{fokkema:15}. We remark that with these boundary conditions, supersymmetric ground states occur for even $L$.

\section{Spin-full CFT with non-linear ${\cal N}=4$ supersymmetry}
\label{sec:N=4spin}
We now turn to the $SU(3)_{k,N}$ CFTs, which underlie the NASS$_k$ quantum Hall states, with filling fraction $\nu={2k \over 2kM+3}$~\cite{ardonne:99}. The simplest fermionic spin-singlet states are the $M=1$ states at filling $\nu={2k \over 2k+3}$, the simplest among those is the $k=1$ state at filling $\nu={2 \over 5}$.

In this section we focus on the CFT underlying the $\nu={2 \over 5}$ spin-singlet state and analyse some of the same issues we considered in the spin-less case. Our main result will be construction of a spinon basis for the finite size spectra of the chiral CFT. This will include a (rather involved) prescription for the $SU(2)$-spin content of multi-spinon states. Our derivation for this combines the results for the spin-less case with the systematics of the spinon basis for $SU(3)_1$ CFT, which were analysed in~\cite{bouwknegt:96}. 

Starting from the spinon basis, we can again do a finitisation. The resulting finite dimensional Hilbert spaces allow a natural interpretation of spin-full lattice fermions satisfying a nearest neighbour exclusion rule, in close analogy with the spin-less case. 

\subsection{Non-linear ${\cal N}=4$ supersymmetry in a $c=2$ CFT}
The CFT underlying the $\nu={2 \over 5}$ spin-singlet quantum Hall state is a $c=2$ theory with scalar fields $\varphi_c$, $\varphi_s$ describing the charge and spin degree of freedom, as measured by the $U(1)$ charge and $SU(2)$ spin currents
\begin{equation}
  J^C =  i\sqrt{2 \over 5} \partial \varphi_c, \qquad J^{\pp,\mm} = e^{\pm i \sqrt{2}\varphi_s}, \quad J^0 = \frac{i}{ \sqrt{2}} \partial \varphi_s .
\end{equation}
The spin currents $J^a(z)$ give rise to an $SU(2)_1$ affine Kac-Moody algebra, with commutators
\begin{equation}
  [ J^a_m,J^b_n ]= m\ d^{ab}\delta_{m+n} + f^{ab}{}_{c} J^c_{m+n}.
\end{equation}
The adjoint index $a$ takes values $a=\pp,0,\mm$. The metric is $d^{\pp\mm}=1, d_{\pp\mm}=1$, $d^{00}=\frac{1}{2}$, $d_{00}=2$ and the structure constants follow from $f^{\pp\mm}{}_0=2$.

As for the spin-less case, the operators describing the creation and annihilation of the fundamental fermion derive from dimension ${3 \over 2}$ currents in the CFT,
\begin{equation}
  G_{\uparrow, \downarrow} = \sqrt{\frac{4}{3}} e^{i\sqrt{\frac{5}{2}} \varphi_c \mp i \sqrt{\frac{1}{2}} \varphi_S},
  \quad
  \bar{G}^{\uparrow, \downarrow} = \sqrt{\frac{4}{3}}  e^{-i\sqrt{\frac{5}{2}} \varphi_c \pm i \sqrt{\frac{1}{2}} \varphi_S}.
\end{equation}
These currents carry both spin and charge, and they constitute a form of ${\cal N}=4$ superconformal symmetry. The non-vanishing (anti-)commutators among the modes of $J^C$, $J^a$,
$G_{\alpha}$ and $\bar{G}^\alpha$ are 
\begin{subequations}
\begin{align}
 [J_m^C, G_{\alpha, n}] &= G_{\alpha, m+n} \\
  [J_m^C, \bar{G}^\alpha_n ] &=-\bar{G}^\alpha_{m+n}\\
 [J^a_m, G_{\alpha, n}]&= (t^a)^\beta_\alpha G_{\beta, m+n} \\
 [J^a_m, \bar{G}^\alpha_n] &= (t^a)^\alpha_\beta \bar{G}^\beta_{m+n}
 \end{align}
 \begin{equation}
 \begin{aligned}
\{ \bar{G}^\alpha_r, G_{\beta,s} \} &=  \frac{2}{3} (r^2-\frac{1}{4}) \delta^\alpha_\beta \delta_{r+s} - \frac{2}{3} (t_a)^\alpha{}_\beta J^a_{r+s} (r-s)  \\
&- \frac{5}{3} \delta^\alpha_\beta J^C_{r+s}(r-s)  +\frac{10}{3} (t_a)^\alpha{}_\beta (J^C J^a)_{r+s} \\
 & + \delta^\alpha_\beta \left[ \frac{10}{3} (L^C)_{r+s} + \frac{2}{3} (L^S)_{r+s}\right].
 \end{aligned}
\end{equation}
\end{subequations}
The spinor indices $\alpha,\beta$ take values $+,-$ and we have $(t^{\pp})^-{}_+=(t^{\mm})^+{}_-=1$, $(t^3)^\pm{}_\pm=\pm \frac{1}{2}$.

As compared to the more standard superconformal algebras, the present algebra has some non-standard features, akin to similar structure in all but the simplest ${\cal W}$-algebras. One such feature is the appearance of modes of the field product $(J^CJ^a)(z)$, the other is the appearance of a non-standard combination of charge and spin stress energy tensors $L^C$ and $L^S$. These derive from the following terms in the OPE $\bar{G}^\alpha(z)G_\beta(w)$
\begin{equation} 
  {\delta^\alpha_\beta \over (z-w)} \left[ 2(T^C(w)+T^S(w))+{4 \over 3}(T^C(w)-T^S(w)) \right] \ .
\end{equation}
The term $2T(z)=2(T^C(w)+T^S(w))$ is a standard OPE descendant term, while ${4 \over 3}(T^C(w)-T^S(w))$ constitutes an independent field that is primary with respect to $T(z)$.

By analogy with the spin-less case we can now define four supercharges
\begin{equation}
  Q_{\alpha}=G_{\alpha,-1/2}, \quad \bar{Q}^\alpha =\bar{G}^\alpha_{1/2}.
\end{equation}
They satisfy the algebra
\begin{equation}
\begin{aligned}
  \{ \bar{Q}^\alpha, Q_\beta \} &=  - \frac{2}{3} (t_a)^\alpha{}_\beta J^a_0 - \frac{5}{3} \delta^\alpha_\beta J^C_0 + \frac{10}{3} (t_a)^\alpha{}_\beta (J^C J^a)_0  \\
  &+ \delta^\alpha_\beta \left[ \frac{10}{3} (L^C)_0+ \frac{2}{3} (L^S)_0\right].
  \end{aligned}
\end{equation}
The complete algebra generated by the zero-modes $J^a_0$, $J^C_0$ of the $SU(2)\times U(1)$ algebra, combined with the supercharges $Q_{\alpha}$, $\bar{Q}^\alpha$, is reminiscent of the Lie superalgebra $SU(2|1)$, but is essentially more complicated due to the presence of the zero-mode of the current product $(J^CJ^a)(z)$ and the terms proportional to $L^C_0$ and $L^S_0$. 

Taking a trace, one may identify a Hamiltonian $H_1$ as
\begin{equation}
  H_1 = \{ \bar{Q}^\alpha, Q_\alpha \} =  - \frac{10}{3} J^C_0 +  \frac{20}{3} (L^C)_0+ \frac{4}{3} (L^S)_0 .
  \label{eq:H1spin}
\end{equation}
We note that it is not $H_1$ but rather the operator 
\begin{equation}
  H^\prime_1 = L^C_0-{1 \over 2} J^C_0
  \label{eq:H1primespin}
\end{equation}
that commutes with the supercharges,
\begin{equation}
  [ H_1, Q_\alpha ] \neq 0, \quad [ H_1, \bar{Q}^\alpha]\neq 0, \qquad [H^\prime_1, Q_\alpha]=0, \quad [H^\prime_1, \bar{Q}^\alpha]=0 .
\end{equation}

\subsection{Spinon multiplets}

General chiral vertex operators in our 2-scalar CFT take the form
\begin{equation}
\label{eq:vqs}
  V^{q,s} = e^{i \frac{q}{\sqrt{10}} \varphi_c \pm i \frac{s}{\sqrt{2}} \varphi_s}, \qquad h=\frac{q^2}{20} +\frac{s^2}{4}.
\end{equation}
We define the basic spinon fields $(\phi^\alpha,\tilde{\phi})$ of conformal dimensions $({3 \over 10}$, ${4 \over 5})$, as 
\begin{equation}
  \phi^{\alpha} = e^{-i \frac{1}{\sqrt{10}} \varphi_c  \pm i \frac{1}{\sqrt{2}} \varphi_s}, 
  \quad \tilde{\phi} = e^{i \frac{4}{\sqrt{10}} \varphi_c}
\end{equation}
with $\alpha = \uparrow, \downarrow$. The supercurrents act as
\begin{equation}
\begin{aligned}
   [ G_{\alpha, s-1/2}, \phi^\beta_{-{3 \over 10}-n} ] &= \sqrt{4 \over 3} \delta_\alpha^\beta \tilde{\phi}_{-{4 \over 5}+s-n}\\ 
   \{ \bar{G}^\alpha_{s+1/2}, \tilde{\phi}_{-{4 \over 5}-n} \} &= \sqrt{4 \over 3}({5 \over 3}n-{2 \over 3}s+1)\phi^\alpha_{-{3 \over 10}+s-n}.
\end{aligned}
\end{equation}
A second basic spinon multiplet is formed by $(\chi,\tilde{\chi}_\alpha,\chi^a)$ of conformal dimensions $({1 \over 5}$, ${7 \over 10}$, ${6 \over 5}$), 
\begin{equation}
\begin{aligned}
  &\chi= e^{-2\frac{i}{\sqrt{10}} \varphi_c}, 
  \qquad 
 && \tilde{\chi}_{\alpha} = e^{i\frac{3}{\sqrt{10}} \varphi_c  \mp \frac{i}{\sqrt{2}} \varphi_s} \\
 & \chi^{\pp,\mm} = e^{i \frac{3}{\sqrt{10}} \varphi_c\pm 2\frac{i}{\sqrt{2}} \varphi_s},
  \qquad &&\chi^3=\partial \varphi_s e^{i \frac{3}{\sqrt{10}} \varphi_c}\ .
\end{aligned}
\end{equation}
On these the supercurrents act as
\begin{equation}
\begin{aligned}
   [ G_{\alpha, s-1/2}, \chi_{-{1 \over 5}-n} ] =& \sqrt{4 \over 3} \tilde{\chi}_{\alpha,-{7 \over 10}+s-n}\\
   \{ \bar{G}^\alpha_{s+1/2}, \tilde{\chi}_{\beta,-{7 \over 10}-n} \} =& \sqrt{4 \over 3} \left[ \delta^\alpha_\beta({5 \over 2}n-{3 \over 2}s+1)\chi_{-{1 \over 5}+s-n}\right.
    \\
    & \left.+ (t_a)^\alpha{}_\beta \chi^a_{-{1 \over 5}+s-n} \right].
\end{aligned}
\end{equation}
We remark that the algebraic structure displayed here can understood by keeping in mind that this is a deformation (set by $M=1$) of the structure of the $SU(3)_1$ current algebra. The latter has 8 currents transforming in the adjoint representation of $SU(3)$; of these, $(J^a,J^C)$ form the reduced bosonic symmetry $SU(2)_1\times U(1)$, while the others are deformed into supercurrents $\bar{G}^\alpha$, $G_\alpha$. In the $SU(3)_1$ theory the basic multiplets are the triplet (the representation $\mathbf{3}$ of $SU(3)$) and anti-triplet (representation $\mathbf{\bar{3}}$); these deform into the multiplets ($\phi^\alpha$, $\tilde{\phi}$) and ($\chi$, $\tilde{\chi}_\alpha$). Note however that the latter multiplet, under the action of the four supercharges, combines with the $\chi^a$ to form a larger $(4_B+2_F)$ dimensional multiplet. 

\subsection{Multi-spinon basis}
Our choice of reference state for the construction of a spinon basis is the state $|\bar{\chi}\rangle$, of charge ${2 \over 5}$ and conformal dimension $\frac{1}{5}$, created from the CFT vacuum by the lowest mode of the vertex operator with $(q=2,s=0)$. We note that the following three states are annihilated by all four supercharges, and thus have $H_1=0$,
\begin{equation}
  |0\rangle = \epsilon_{\alpha\beta} \phi^\alpha_{3 \over 10} \phi^\beta_{-{1 \over 10}} |\bar{\chi}\rangle, 
  \quad
  |\bar{\phi}^\alpha\rangle = \phi^\alpha_{-{1 \over 10}} |\bar{\chi}\rangle, 
  \quad  
  |\bar{\chi}\rangle \ .
  \label{eq:spinsusysinglets}
\end{equation}
Again by analogy with the spin-less case we define a general multi-spinon state as a `word' of  $N$ $\phi^{\alpha}$,
\begin{equation}
  \phi^{\alpha_N}_{\frac{(4N-5)}{10} -n_N} \ldots \phi^{\alpha_2}_{\frac{3}{10} -n_2} \phi^{\alpha_1}_{-\frac{1}{10} -n_1} |\bar{\chi}\rangle.
\end{equation}
This state has $J^C_0 = \frac{2-N}{5}$ and conformal dimension
\begin{equation}
h=-\frac{1}{10} (2N^2-3N) + \sum_{j=1}^N n_j + \frac{1}{5}.
\end{equation}
We can define further states by replacing any of the $\phi^{\alpha_j}_{\frac{4j-5}{10}-n_j}$ by the corresponding superpartner $\tilde{\phi}_{\frac{4j-10}{10}-n_j}$. For a generic choice of the $n_j$, the resulting set contains $3^N$ independent states in the chiral CFT Hilbert space, with $SU(2)$ spin content given by $[{1 \over 2}\oplus 0]^{\otimes N}$. However, as for the spin-less case, many of these states vanish when the $n_j$ are sufficiently close together. 

\begin{figure}
\centering
  \includegraphics{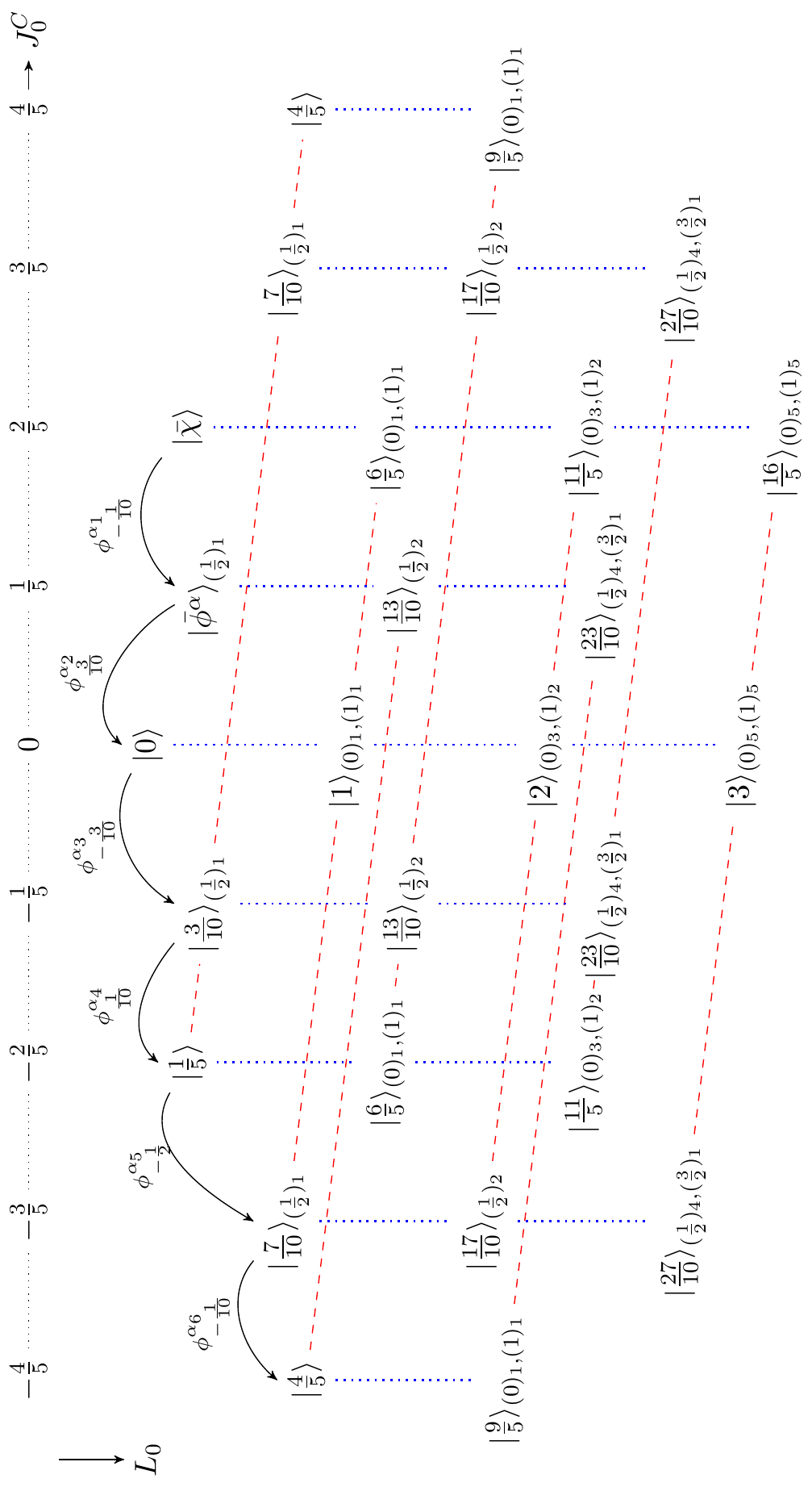}
  \caption{Chiral spectrum of the non-linear ${\cal N}=4$ supersymmetric CFT at $c=2$. The supercharges $Q_\alpha$ and $\bar{Q}^\alpha$ act parallel to the red dashed lines. The spin content of the states are denoted by $(s)_j$, meaning that the spin content is $s$ with multiplicity $j$. \label{fig:cftspinsusy}}
\end{figure}

It is a rather involved problem to specify the precise number of states (as well as the $SU(2)$-spin content) for a given choice of the $\{n_j\}$. To determine these rules, we combined the insights for the spin-less case with the results, presented in~\cite{bouwknegt:96}, for the $SU(3)_1$ model. For the latter, one builds a spinon basis by forming multi-$\phi^A$ words, with $\phi^A$ the fundamental spinon in the representation $\mathbf{3}$ of $SU(3)$. The `Pauli-principle' for these $SU(3)$ spinons involves a combination of (anti\mbox{-})symmetrisations and trace-subtractions, the latter specified by so-called hook rules~\cite{bouwknegt:96}. The complete recipe for the present ${\cal N}=4$ supersymmetric case, which we detail in appendix~\ref{sec:AppC}, takes a very similar form.

Let us first explore the `extreme choices' for a sequence $\{n_j\}$. It turns out that at most two consecutive $n_j$ can be equal; if that happens the corresponding spin-indices are anti-symmetrized. The sequence with the lowest possible $n_j$ is thus (assuming an even number of spinons)
\begin{equation}
  \{j, j, \ldots, 1, 1, 0, 0\}: [0,{1 \over 2}].
\end{equation}
This notation indicates that the spin-content of the leading state, at $J^C_0=-{2j \over 5}$, $h=\frac{j^2}{5}$ is a singlet, $s=0$, while there is a single superpartner at $J^C_0={5-2j \over 5}$, which is a doublet under $SU(2)$-spin, $s={1 \over 2}$. If $j=0$ the superpartner is absent (we already indicated that this is the CFT vacuum state, which is a singlet under the supersymmetry). These states can up to $j=2$ be seen in Figure~\ref{fig:cftspinsusy}.

The lowest sequence $\{n_j\}$ leading to the full count of $3^N$ states is 
\begin{equation}
\begin{aligned}
 \{2N-1, \ldots 5, 3, 1\}: \Big[ &\left(\frac{1}{2}\right)^{\otimes N}, \left(\left(\frac{1}{2}\right)^{\otimes N-1}\right)_N,  \left(\left(\frac{1}{2}\right)^{\otimes N-2}\right)_{\tbinom{N}{2}},  \\
 &  \ldots, \left(\left(\frac{1}{2}\right)^{\otimes 1}\right)_{\tbinom{N}{N-1} }, 0 \Big],
 \end{aligned}
\end{equation}
where $(\left(\tfrac{1}{2}\right)^{\otimes k})_j$ denotes the tensor product of $k$ times spin $\tfrac{1}{2}$ with multiplicity $j$.
This state with all $\phi^\alpha$ replaced by $\tilde{\phi}$ has the form
\begin{equation}
  \tilde{\phi}_{{4N-10 \over 10}-(2N-1)} \ldots \tilde{\phi}_{-{2 \over 10}-3}\tilde{\phi}_{-{6 \over 10}-1}  |\bar{\chi}\rangle,
\end{equation}
and it precisely represents the leading (lowest $L_0$) state in the CFT sector with $U(1)$ charge $J^C_0={2+4N \over 5}$ and $h=\frac{1}{5} (2N+1)^2$. 

A general 1-spinon state has spin content $[{1 \over 2},0]$ if $n_1>0$ while for $n_1=0$ this is the $s={1 \over 2}$ state $|\bar{\phi}^\alpha\rangle$, which is a supersymmetry singlet. For 2-spinon states we have 
\begin{equation*}
\begin{aligned}
  n_2&=n_1=0,  && [0,-,-]\\
   n_2&=n_1>0, && [0,\tfrac{1}{2},-]\\               
   n_2&>n_1, n_1=0, && [0+1,\tfrac{1}{2},-]\\                     
   n_2&=n_1+1, n_1>0, && [0+1,(\tfrac{1}{2})_2,-]\\                               
   n_2&>n_1+1, n_1>0, && [0+1,(\tfrac{1}{2})_2,0],
\end{aligned}
\end{equation*}
where $-$ means that there is no state.

We refer to appendix~\ref{sec:AppC} for a precise (but conjectured) prescription of the spin-content corresponding to general sequence $\{n_j\}$. Our analysis leading to these rules was in a large part based on the requirement that the multi-spinon states form a basis for the chiral CFT space of states. We have refrained from constructing operators such as $H_2$, which permit a sharper formulation of the independence and completeness of the multi-spin basis, as we demonstrated in the case without spin. We have made the important check that our results for the exclusion rules do lead to the correct multiplicities and spin content in the CFT modules, up to high values of $L_0$. In addition, we established that the finitisation of this same multi-spinon basis leads to highly structured finite-$L$ partition sums, which permit a natural interpretation of partition sums of spin-full lattice fermions with a simple and spin-independent nearest neighbour exclusion rule. We present this result in our next subsection.

\subsection{Finitisation and CFT characters}
\label{sec:susyspinfinite}
We can finitise the CFT spectrum by imposing the constraint that the highest spinon mode-index is of the form
\begin{equation}
  n_N -{4N -5 \over 10} = {2L-3 \over 10} - l,
  \label{eq:finspin}
\end{equation}
with $l$ a non-negative integer. We define the corresponding partition function as
\begin{equation}
  Q_L(w,z,q) = \tr [w^{5J^C_0} z^{2J^0_0} q^{L_0-{1 \over 2}J_0^C}] \ .
\end{equation}
We discuss some small sizes in Table~\ref{tab:finitespin2}.
\begin{table}
\centering
  \begin{tabular}{lrl}\toprule
  Length & $n_j$ &  spin content\\
  \midrule
    $L=0$ & $00$ & $[0]$ \\[4pt]
    $L=1$ & $1100$ & $ [0,\frac{1}{2}] $\\[4pt]
    $L=2$ & $221100$ & $ [0,\frac{1}{2}]$\\
             & $ 0 $ & $ [-,\frac{1}{2}]   $\\[4pt]
    $L=3$  & $ 33221100 $ & $ [0,\frac{1}{2},-]$\\
             & $ 110 $ & $ [-,\frac{1}{2},1]  $ \\
             & $ 100 $ & $  [-,\frac{1}{2},0]$ \\[4pt]
    $L=4$  & $ 4433221100$ & $ [0,\frac{1}{2},-] $\\
             & $ 22110 $ & $ [-,\frac{1}{2},1] $ \\
             & $ 22100 $ & $ [-,\frac{1}{2},0+1] $ \\
             & $ 21100 $ & $ [-,\frac{1}{2},0+1] $\\
             & $ -  $ & $  [-,-,0]$ \\[4pt]
    $L=5$   & $ 554433221100$ & $ [0,\frac{1}{2},-,-]$ \\
             & $ 3322110 $ & $ [-,\frac{1}{2},1,-]$ \\
             & $ 3322100 $ & $ [-,\frac{1}{2},0+1,-] $\\
             & $ 3321100 $ & $ [-,\frac{1}{2},0+1_2,\frac{3}{2}]$  \\
             & $ 3221100 $ & $ [-,\frac{1}{2},0+1,-]$  \\
             & $ 11 $ & $ [-,-,0,\frac{1}{2}]$\\
             & $ 10 $ & $ [-,-,0+1,\frac{1}{2}]$\\
             & $ 00 $ & $ [-,-,0,-]$\\
            \bottomrule
  \end{tabular}
  \caption{Multi-spinon states and spin content that contribute to finitised partition functions $Q_L$ in the spin-full CFT with non-linear $\mathcal{N}=4$ supersymmetry.}
  \label{tab:finitespin2}
\end{table}
These translate into partition sums
\begin{align*}
  Q_0 = & 1\\
   Q_1= & w^{-2}q^{2 \over 5}\left[1+w^5(z+z^{-1})\right]\\
   Q_2= & w^{-4}q^{6 \over 5}\left[ 1+ w^5(z+z^{-1})(1+q^{-1}) \right]\\
   Q_3=& w^{-6}q^{12 \over 5}\left[1+w^5(z+z^{-1})(1+q^{-1}+q^{-2})+ w^{10} [(z^{-2}+1+z^2)q^{-1}+q^{-2}]\right]\\
   Q_4= & w^{-8}q^{4}\left[1+w^5(z+z^{-1})(1+q^{-1}+q^{-2}+q^{-3})\right.\\
  & \qquad \qquad \left. + w^{10} [(z^2+1+z^{-2})+q^{-1}](q^{-1}+q^{-2}+q^{-3}) \right]\\
   Q_5=& w^{-10} q^{6} \left[ 1+w^5(z+z^{-1})(1+q^{-1}+q^{-2}+q^{-3}+q^{-4}) \right.\\
  &  \qquad\qquad \left. + w^{10} [(z^2+1+z^{-2})+q^{-1}](q^{-1}+q^{-2}+2q^{-3}+q^{-4}+q^{-5}) \right.\\
  & \qquad\qquad \left. + w^{15}[ (z^3+z^1+z^{-1}+z^{-3})q^{-3} + (z^{-1}+z)(q^{-4}+q^{-5}) ]\right].
\end{align*}
The general formula for $Q_L$ is  in essence the same as in the spin-less case up to an extra factor, which we call $f_p(z,q)$, in every term of the sum,
\begin{equation}
  Q_L(w,z,q)=\sum_{p=0,1,2,\ldots}^{\left \lfloor{(L+1)/2}\right \rfloor} w^{5p-2L} q^{\frac{(L-{5p \over 2})(L+1-{5p \over 2})}{5}+\frac{p^2}{4}} f_p(z,q) \binom{L+1-p}{p}_q ,
\end{equation}
with for $p$ even
\begin{equation}
  f_p(z,q)= \binom{p}{p/2}_{q^{-1}}+\sum_{k=0}^{p/2-1} \binom{p}{k}_{q^{-1}} \left(z^{p-2k} + z^{-(p-2k)}\right)
\end{equation}
and for $p$ odd
\begin{equation}
  f_p(z,q) =\sum_{k=0}^{(p-1)/2} \binom{p}{k}_{q^{-1}} \left(z^{p-2k} + z^{-(p-2k)}\right)\ . 
\end{equation}
In the limit $L\to\infty$ these expressions lead to 
\begin{equation}
  \underset{\lim l \rightarrow \infty} {Q_{5l+k}}(w,z,q) 
  =\underset{s \in 2 \mathbb{Z}+1\ \text{for}\  m \ \text{odd}}{ \underset{s \in 2 \mathbb{Z}\ \text{for}\  m \ \text{even}}{\sum_{m \equiv -2k \ \text{mod}\ 5}}} z^s w^m q^{\frac{m^2}{20} + \frac{s^2}{4}-\frac{m}{10}} \left( \prod_l \frac{1}{1-q^l}\right)^2,
\end{equation}
where $m$ corresponds to the $q$ in equation~\eqref{eq:vqs} and we recognise the characters of the CFT modules that build the $c=2$ CFT.

\subsection{Spin-full lattice models}
Putting $q=1$ the partition sums $Q_L(w,z)$ agree with the partition sum for a lattice model with spin ${1 \over 2}$ fermions on an $L$ site open chain, subject to the constraint that each site has at most one fermion, and that nearest neighbour sites cannot be simultaneously occupied. In this correspondence the number of particles $f$ on the chain is given by
\begin{equation}
  f = J_0^C+{2L \over 5}.
  \label{eq:fj0c}
\end{equation}
For the example $L=3$ we get for $q=1$, $Q_3=w^{-6} + 3 w^{-1} (z + z^{-1}) + w^{4} (z^{-2} + 2 + z^2)$, where the power of $w$ is equal to $5 J_0^C$. Using eq.~\eqref{eq:fj0c} this corresponds to the allowed configurations 
\begin{eqnarray}
  f=0 & & \circ \, \circ \, \circ 
          \nonu 
          f=1 & & \alpha_1 \, \circ \, \circ,  \  \circ  \, \alpha_2 \, \circ, \  \circ\, \circ\, \alpha_3
                  \nonu
                  f=2 & & \alpha_1 \, \circ \, \alpha_3  ,
\end{eqnarray} 
where $\alpha_j$ denotes a spin $\frac{1}{2}$ fermion.
We remark that this configuration space differs from the one considered in an alternative approach to spin-full supersymmetric lattice models proposed in~\cite{santachiara:05}.

Following the logic we pursued for the spin-less case, one may try to define ${\cal N}=4$ supercharges $Q_\alpha$ and $\bar{Q}^\alpha$ directly on the lattice and define a Hamiltonian $H_1$ through a formula like equation~\eqref{eq:H1spin}. If enough of the algebraic structure of the CFT persists on the lattice, one would expect lattice model ground states in correspondence to the three CFT ground states~\eqref{eq:spinsusysinglets}. Tracing through the finitisation, these would correspond to $f$-particle states with 
\begin{equation}
\begin{aligned}
  & |\bar{\chi}\rangle: && L=5l-1,\quad &&f=2l\\
  & |\bar{\phi}^\alpha\rangle: && L=5l+2,\quad &&f=l+1\\
  & |0\rangle: && L=5l,\quad &&f=2l .
  \end{aligned}
\end{equation}
A naive definition in terms of dressed creation and annihilation operators (as in equation~\eqref{eq:dressedc}),
\begin{equation}
  Q_\alpha = d_\alpha, \qquad \bar{Q}^\alpha = (d_\alpha)^\dagger
\end{equation}
does give the expected supersymmetric ground state for $L=2$ but not for $L=4,5$ and higher. We have observed that modified definitions of the supercharges can be tuned such as to produce supersymmetric ground states on small-sized lattices, suggesting an iterative procedure to define these charges \cite{FS:progress}.

The algebraic structure of the spin-full case is considerably more complicated than that of the spin-less case, both in the CFT and on the lattice. For one thing, the notion of a Witten index, which protects supersymmetric ground states in the spin-less case, is missing in the spin-full case. In addition, in the spin-full case the Hamiltonian $H_1$, defined through equation~\eqref{eq:H1spin} fails to commute with the supercharges. In the CFT, the operator $H_1^\prime$ that does commute with the supercharges, see equation~\eqref{eq:H1primespin}, is the sum of a multiple of $H_1$ and a term which is expressed in $L_0^S$ and thus originates in the spin sector of the model. This observation may hold a clue for an improved definition of a lattice model Hamiltonian that commutes with appropriately defined supercharges and that leads to a lattice model mimicking some of the algebraic structure of the CFT. 

\bigskip
\noindent
{\bf Acknowledgements.}\ We thank Jasper Stokman for discussions. Part of this work was done at the Rudolf Peierls Centre at the University of Oxford and at the Galileo Galilei Institute in Firenze. T.F. is supported by the Netherlands Organisation for Scientific Research (NWO). This research is part of the Delta ITP consortium, a program of NWO that is funded by the Dutch Ministry of Education, Culture and Science (OCW). 

\appendix
\section{Normal ordered field products from BPZ expansions}
\label{sec:appA}
In evaluating the action of operators $Q_1^+$ ad $H_2$ on multi-spinon states, we make extensive use of manipulations involving normal ordered field products in CFT. We here present some of the relations needed to process such products (see for example \cite{bais:87}). We start from two chiral fields (CVO's) with defining OPE
\bea
\lefteqn{\phi^1(z)\phi^2(w) = (z-w)^{\Delta^\prime-\Delta_1-\Delta_2}}
\nonu
&& \times \left[ \phi^\prime(w) + a_1 \partial \phi^\prime(w) (z-w) \right.
\nonu 
&& 
      \left. +\left[  (a_{11}-{3 \over 2(2\Delta^\prime+1)} a_2) \partial^2 \phi^\prime(w) + a_2 (T\phi^\prime)(w) \right] (z-w)^2 + \ldots\right]
\eea
with
\bea
&& a_1={\Delta_1-\Delta_2+\Delta^\prime \over 2 \Delta^\prime},
\quad
a_{11}={(\Delta_1+\Delta_2+\Delta^\prime +1)(\Delta_1-\Delta_2+\Delta^\prime) \over 4\Delta^\prime(2\Delta^\prime+1)}
\nonu
&& a_2={\Delta^\prime(\Delta^\prime-1)+(\Delta_1+\Delta_2)(2\Delta^\prime +1) - 3(\Delta_1-\Delta_2)^2 \over c(2\Delta^\prime+1)+2\Delta^\prime(8\Delta^\prime-5)}.
\eea        
The normal ordered product is defined as 
\be
(\phi^1\phi^2)(w)=\oint{dz \over 2\pi i} {\phi^1(z)\phi^2(w) \over z-w}.
\ee
Using the standard expansion
\be
\phi(z) = \sum_m \phi_m z^{-m-\Delta}
\ee
this leads to
\be
(\phi^1\phi^2)_{-\Delta_1-\Delta_2-m} = \sum_{l\geq 0} \phi^1_{-\Delta_1-l} \phi^2_{-\Delta_2-m+l} \pm \sum_{l>0} \phi^2_{-\Delta_2-m-l}  \phi^1_{-\Delta_1+l}.
\ee
with the relative sign equal to minus if both fields are fermionic in nature.

In our computations in the ${\cal N}=2$, $k=1$ CFT we used the explicit normal ordered field products
\bea
&& (J\phi^-) = - \partial \phi^-, \qquad  (\partial J \phi^-) = 2(T\phi^-)-3\partial^2\phi^-
\nonu
&&  (G^+\phi^-) = \sqrt{2 \over 3} {3 \over 2}(\partial \phi^+), \qquad (\partial G^+ \phi^-) =  \sqrt{2 \over 3} \left( - {3 \over 5} (T\phi^+) + {27 \over 20} \partial^2 \phi^+ \right)
\nonu
&& (G^-\phi^+) = \sqrt{2 \over 3} \left( 6 (T\phi^-)- {9 \over 2} \partial^2 \phi^- \right).
\eea

\section{Explicit results for two-spinon states}\label{sec:appB}
We present the action of $Q_1^+$ and $H_2$ on 2-spinon states, $\Phi^{\alpha, \beta}_{n_2, n_1}$, and give explicit expressions for the 2-spinon $H_2$ eigenstates,  $\Psi^{\alpha, \beta}_{n_2, n_1}$. 

\subsection{Action of $Q_1^+$ and $H_2$ on 2-spinon states}
To streamline notation we write
\begin{equation}
  \begin{aligned}
    & \Phi^{-,-}_{n_2,n_1} =\phi^-_{-1/6-n_2} \phi^-_{1/6-n_1} |+\rangle, \quad  &&\Phi^{+,-}_{n_2,n_1} =\phi^{+}_{-2/3-n_2} \phi^-_{1/6-n_1} |+\rangle\\
    & \Phi^{-,+}_{n_2,n_1} =\phi^-_{-1/6-n_2} \phi^{+}_{-1/3-n_1} |+\rangle, \quad &&\Phi^{+,+}_{n_2,n_1} =\phi^{+}_{-2/3-n_2} \phi^{+}_{-1/3-n_1} |+\rangle.
  \end{aligned}
\end{equation}
We obtain
\begin{equation}
  \begin{aligned}
    \sqrt{3 \over 2} \, Q_1^+ \, \Phi^{-,-}_{n_2,n_1} &= (n_2+{1 \over 3}) \Phi^{+,-}_{n_2,n_1} + (n_1-{2 \over 3}) \Phi^{-,+}_{n_2,n_1} 
    \\ &\quad
    + {2 \over 3} \sum_{l>0}  \Phi^{+,-}_{n_2+l,n_1-l}  - {2 \over 3} \sum_{l>0} \Phi^{-,+}_{n_2+l,n_1-l} 
    \\
    \sqrt{3 \over 2} \, Q_1^+ \, \Phi^{+,-}_{n_2,n_1} &= -(n_1-{2 \over 3}) \Phi^{+,+}_{n_2,n_1}  - {4 \over 3} \sum_{l>0} \Phi^{+,+}_{n_2+l,n_1-l} 
    \\
    \sqrt{3 \over 2} \, Q_1^+ \, \Phi^{-,+}_{n_2,n_1} &= (n_2-{5 \over 3}) \Phi^{+,+}_{n_2,n_1}  - {4 \over 3} \sum_{l>0}  \Phi^{+,+}_{n_2+l,n_1-l}  
    \\
    \sqrt{3 \over 2} \, Q_1^+ \, \Phi^{+,+}_{n_2,n_1} & = 0 
  \end{aligned}
\end{equation}
and
\begin{equation}
  \begin{aligned}
    H_2 \Phi^{-,-}_{n_2,n_1} &= h_2[n_2,n_1] \Phi^{-,-}_{n_2,n_1} + {4 \over 3} \sum_{l>0}  (n_2-n_1+2l+{1 \over 3}) \Phi^{-,-}_{n_2+l,n_1-l}  
    \\
    H_2 \Phi^{+,-}_{n_2,n_1} &= h_2[n_2,n_1]  \Phi^{+,-}_{n_2,n_1}  
    \\ &\quad
    - {4 \over 3} \sum_{l>0} (3n_2+2l+1)\Phi^{-,+}_{n_2+l,n_1-l}  + {4 \over 3} \sum_{l>0} (2n_1+n_2-2l+{1 \over 3})\Phi^{+,-}_{n_2+l,n_1-l} 
    \\
    H_2 \Phi^{-,+}_{n_2,n_1} & = h'_2[n_2,n_1] \Phi^{-,+}_{n_2,n_1} +  4 n_1 \Phi^{+,-}_{n_2,n_1}  
    \\ &\quad
    - {4 \over 3} \sum_{l>0}  (2n_2+n_1+2l+{2 \over 3})  \Phi^{-,+}_{n_2+l,n_1-l}   + {4 \over 3} \sum_{l>0}  (3n_1-2l)  \Phi^{+,-}_{n_2+l,n_1-l}  
    \\
    H_2 \Phi^{+,+}_{n_2,n_1} &= h_2'[n_2,n_1] \Phi^{+,+}_{n_2,n_1} - {8 \over 3} \sum_{l>0}  (n_2-n_1+{1 \over 3}) \Phi^{+,+}_{n_2+l,n_1-l} .  
  \end{aligned}
\end{equation}
                                            
\subsection{The doublet $[n_2,n_1;-]$: $\Psi_{n_2,n_1}^{-,-}$ and $\Psi_{n_2,n_1}^{+,-}$}
These $H_2$ eigenstates, with eigenvalue
\begin{equation}
  h_2[n_2,n_1] = 2(n_2+{1 \over 3})^2 + 2n_1(n_1-{2 \over 3}),
\end{equation}
take the explicit form
\begin{equation}
  \begin{aligned}
    \Psi^{-,-}_{n_2,n_1} &= \Phi^{--}_{n_2,n_1} + \sum_{k=1}^{n_1} \alpha_k \ \Phi_{n_2+k, n_1-k}
    \\
    \Psi^{+,-}_{n_2,n_1} &= \Phi^{+,-}_{n_2,n_1} + \sum_{k=1}^{n_1} \beta_k \Phi^{+,-}_{n_2+k,n_1-k} +\sum_{k=1}^{n_1-1} \gamma_k \Phi^{-,+}_{n_2 + k, n_1-k},
  \end{aligned}
\end{equation}
with
\begin{equation}
  \begin{aligned}
    \alpha_k &= - \frac{n+2k+\frac{1}{3}}{3} \frac{\left(\frac{2}{3}\right)^{k-1} }{\left(1\right)^k}    \frac{ \left(n+\frac{4}{3}\right)^{k-1}}{\left( n+\frac{5}{3} \right)^k}
    \\
    \beta_k &= \frac{3n_1 -3k-1}{3n_1-1} \frac{\left(\frac{2}{3}\right)^k }{\left(1\right)^k}   \frac{ \left(n+\frac{4}{3}\right)^k}{\left( n+\frac{5}{3} \right)^k}
    \\
    \gamma_{k+1} &= -3 \frac{n_2 + k+1}{3n_1-1} \frac{\left(\frac{2}{3}\right)^{k-1} }{\left(1\right)^{k-1}}   \frac{ \left(n+\frac{4}{3}\right)^{k-1}}{\left( n+\frac{5}{3} \right)^{k-1}},
  \end{aligned}
\end{equation}
where $n=n_2-n_1$.  Our notation employs the Pochhammer symbol,
\be
(a)^k = \prod_{j=1}^k (a+j-1)\ .
\ee

\subsection{The doublet $[n_2,n_1;+]$: $\Psi_{n_2,n_1}^{-,+}$ and $\Psi_{n_2,n_1}^{+,+}$}
These states, with eigenvalue
\begin{equation}
  h^\prime_2[n_2,n_1] = 2(n_2+{1 \over 3})(n_2-{5 \over 3}) + 2n_1(n_1-{2 \over 3}),
\end{equation}
take the form
\begin{equation}
  \label{eq:psialphaplus}
  \begin{aligned}
    \Psi^{-,+}_{n_2,n_1} &= \Phi^{-,+}_{n_2,n_1} + \sum_{k=1}^{n_1-1} \zeta_{k} \ \Phi^{-,+}_{n_2 + k, n_1 -k} + \sum_{k=0}^{n_1} \eta_k \ \Phi^{+,-}_{n_2 + k, n_1 -k}
    \\
    \Psi^{+,+}_{n_2,n_1} &= \Phi^{+,+}_{n_2,n_1} + \sum_{k=1}^{n_1-1} \rho_k \ \Phi^{+,+}_{n_2 +k, n_1-k},
  \end{aligned}
\end{equation}
with
\begin{equation}
  \label{eq:etazetarho}
  \begin{aligned}
    \eta_k &= -3 \frac{n_1-k}{3n_2 +1}  \frac{\left(\frac{2}{3}\right)^k}{(1)^k}  \frac{\left(n+ \frac{1}{3} \right)^k}{\left(n+\frac{2}{3} \right)^k}
    \\
    \zeta_k &= \frac{3n_2 + 3k+1}{3n_2 +1} \frac{\left(\frac{2}{3}\right)^k}{(1)^k}  \frac{\left(n+ \frac{1}{3} \right)^k}{\left(n+\frac{2}{3} \right)^k} 
    \\
    \rho_k &= \frac{\left(\frac{2}{3}\right)^k}{(1)^k}  \frac{\left(n+ \frac{1}{3} \right)^k}{\left(n+\frac{2}{3} \right)^k}.
  \end{aligned}
\end{equation}

\section{Spin-content of spin-full multi-spinon states}
\label{sec:AppC}
We specify our conjectured rule for determining the $SU(2)$ spin content of multi-spinon states for general sequence $n_N\geq n_{N-1}\geq \ldots n_1\geq 0$. In addition to the examples specified below, the main text in section \ref{sec:N=4spin} gives many more examples and discusses the status of this conjecture.

\begin{itemize}[align=left]
\item[\textbf{Step 1.}] As soon as $n_{j_0}>n_{j_0-1}+1$ the result is the free product of contributions from $j\geq j_0$ and $j\leq j_0-1$. 
\end{itemize}
Hence we can restrict our attention to indecomposable patterns, where consecutive $n_j$ differ by $0$ or $1$.
\begin{itemize}[align=left]
\item[\textbf{Step 2.}] At most two consecutive $n_j$ are equal. When this happens the corresponding spinons $\phi^\alpha$ are anti-symmetrized and form the singlet $\chi$. This we represent as a \emph{pair} $nn$.

\item[\textbf{Step 3.}] We delete $00$ and delete repeating pairs of equal $n_j$ from the sequence such that only one pair remains, where all integers to the left of the deletion are shifted down accordingly. Examples
\begin{equation}
  22100 \rightarrow 221, \quad 7766554 \rightarrow 554,  \quad 44322110 \rightarrow 332110, \quad \text{etc.} 
\end{equation}

\item[\textbf{Step 4.}] For a single $s={1 \over 2}$ spinon $n$ with $n>0$ an $s=0$ superpartner $\tilde{n}$ is possible; for an $s=0$  pair $nn$, an $s={1 \over 2}$ superpartner $\widetilde{nn}$ is possible.
\end{itemize}
We are now down to words with letters $n$, $\tilde{n}$, $nn$ and $\widetilde{nn}$ without $00$ or repeating pairs; reading from left to right, a value $n$ is followed by $n-1$.
\begin{itemize}[align=left]
\item[\textbf{Step 5.}] At the top level (no superpartners thus lowest $J_0^C$ eigenvalue), the spin content is the free product of the spin-${1 \over 2}$ associated with a single $n$.

\item[\textbf{Step 6.}] The only allowed 2-letter combinations involving one or two superpartners are, with $m=n-1$,
\begin{equation}
  n\,\widetilde{mm}, \qquad n\,\tilde{m}, \qquad \tilde{n}\,m, \qquad \widetilde{nn}\,m.
\end{equation}
\end{itemize}
Let us pause and consider the 3-spinon case. After following steps 1-6 we have two indecomposable patterns involving one superpartner,
\begin{equation}
  \Psi = (n+1)\, \widetilde{nn}, \qquad  \Phi = \widetilde{nn}\, (n-1) \ .
\end{equation}
Both lead to spin content $0+1$ with the exception of $\widetilde{11}\, 0$ which has the triplet $s=1$ only. The means that we have symmetrized the spin content  of the word $\Phi$ in this particular situation.
\begin{itemize}[align=left]
\item[\textbf{Step 7.}] It turns out that, more generally, certain patterns involving (products of) words $\Psi$ and $\Phi$ need to be symmetrized, depending on the neighbouring letters. 
We denote the symmetrisation by $\underbrace{\text{pattern}}$. The list is
\begin{equation}
\begin{aligned}
  & \tilde{r} \ \underbrace{k \ \Phi \ldots \Phi} \ \tilde{n}\\
  & \underbrace{\Phi \ldots \Phi}\  \tilde{n}, \quad \underbrace{\Phi \ldots \Phi  \ \widetilde{nn} \ m} \ kk, 
     \quad \underbrace{\Phi \ldots \Phi \ 11 \ 0}, \quad \underbrace{\Phi \ldots \Phi  \ \widetilde{nn}}\\
  & \tilde{n} \underbrace{\Psi \ldots \Psi \ 0}, \quad \tilde{n} \ \underbrace{\Psi \ldots \Psi \ m} \ kk,
     \quad \tilde{n} \ \underbrace{\Psi \ldots \Psi}\\
  & \underbrace{\widetilde{nn} \ m} \ kk, \quad \underbrace{\widetilde{11} \ 0}.
\end{aligned}
\end{equation}
The pattern $\Phi \ldots \Phi$ means that $\Phi$ appears multiple times in a row, but the rule also applies to just a single occurrence of $\Phi$.

\item[\textbf{Step 8.}] After steps 1-7 one last reduction is needed. A word of the form $\widetilde{44}32\widetilde{11}$ contributes $\frac{1}{2}^{\otimes 4} - 0$. Our final step takes care of the subtraction of the singlet in configurations where such a pattern occurs. If this pattern exists as part of a word, the case where it forms a singlet, which we denote by a hook, should be deleted. If two of these hooks are possible double hooks should be added, triple hooks deleted again, etc. Some examples of this reduction are 
\end{itemize}
\begin{align*}
\widetilde{44}32\widetilde{11} &= (\tfrac{1}{2})^{\otimes 4} -\underbracket{\widetilde{44}32\widetilde{11}}= (\tfrac{1}{2})^{\otimes 4} - (0)\\
\underbrace{\widetilde{66} 5 \widetilde{44}} 32 \widetilde{11} &=((\tfrac{3}{2}) \otimes (\tfrac{1}{2})^{\otimes 3}) -\underbrace{\widetilde{66} 5} \underbracket{ \widetilde{44}32 \widetilde{11}} =(\tfrac{3}{2}) \otimes (\tfrac{1}{2})^{\otimes 3}-(1)\\
\widetilde{77}65\widetilde{44}32\widetilde{11} &=(\tfrac{1}{2})^{\otimes 7}-\underbracket{\widetilde{77}65\widetilde{44}}32\widetilde{11}-\widetilde{77}65\underbracket{\widetilde{44}32\widetilde{11} } =(\tfrac{1}{2})^{\otimes 7}- (\tfrac{1}{2}^{\otimes 3})-(\tfrac{1}{2})^{\otimes 3}.
\end{align*}

\end{document}